\documentclass[twocolumn]{aastex62}

\usepackage{diagbox}
\usepackage{makecell}

\newcommand\aastex{AAS\TeX}

\newcommand{\tabincell}[2]{\begin{tabular}{@{}#1@{}}#2\end{tabular}}
\newcommand       \Angstrom     {\,{\rm \AA}}

\newcommand       \mum           {\,{\rm \mu m}}

\newcommand{\kms}{\,km\,s$^{-1}$}
\newcommand{\accunit}{$_{\odot}$ yr$^{-1}$}
\received{XXX}
\revised{XXX}
\accepted{XXX}
\submitjournal{ApJs}

\shorttitle{\aastex\ The stellar groups in the field of Taurus}

\begin{document}

\title{The old moving groups in the field of Taurus}

\correspondingauthor{Min Fang}
\email{mfang.cn@gmail.com}

\author{Jiaming Liu}
\affiliation{Key Laboratory of Optical Astronomy, National Astronomical Observatories, Chinese Academy of Sciences, Beijing 100101, China}

\author{Min Fang}
\affiliation{Purple Mountain Observatory, Chinese Academy of Sciences, 10 Yuanhua Road, Nanjing 210023, China}

\author{Hao Tian}
\affiliation{Key Lab of Space Astronomy and Technology, National Astronomical Observatories, CAS, Beijing, 100101, China}

\author{Chao Liu}
\affiliation{Key Lab of Space Astronomy and Technology, National Astronomical Observatories, CAS, Beijing, 100101, China}
\affiliation{University of Chinese Academy of Sciences, Beijing 100049, China}

\author{Chengqun Yang}
\affiliation{Shanghai Astronomical Observatory, 80 Nandan Road, Shanghai 200030, People’s Republic of China}

\author{XiangXiang Xue}
\affiliation{Key Laboratory of Optical Astronomy, National Astronomical Observatories, Chinese Academy of Sciences, Beijing 100101, China}

\begin{abstract}

In this work, we present a systematic search for stellar groups in the Taurus field by applying DBSCAN algorithm to the data from {\it Gaia} DR2. We find 22 groups, consisting of 8 young groups (Groups 1--8) at ages of 2--4\,Myr and distances of $\sim$130--170\,pc, 14 old groups (Groups 9-22) at ages of 8--49\,Myr and distances of $\sim$110--210\,pc.  We characterize the disk properties of group members and find 19 new disk-bearing stars, 8 of which are in the young groups and 11 others belong to the comparatively old groups at ages of 8--11\,Myr. We characterize the accretion properties of the group members with H$\alpha$ emission line in their LAMOST spectra, and discover one source in Group 10 at an age of 10\,Myr which still shows accretion activity. We investigate the kinematic relations among the old groups, and find that Group 9 is kinematically related to the known Taurus members and exclude any kinematic relations between Groups~10-22 and the known Taurus members.
\end{abstract}


\section{introduction}\label{intro}

Taurus is one of the most famous nearby (130--160\,pc, \citealt{Torres2009}) star forming region. This region has been surveyed with various telescopes at different wavelengths, e.g. the infrared imaging survey with the IRAS, Spitzer, and WISE telescopes \citep{Ken1990,Luh2010,Reb2010,Reb2011,Esp2014}, the optical imaging survey \citep{Bri2002,Sle2006}, the UV imaging survey with the GALEX \citep{Fin2010,Gom2015}, the X-ray survey with the XMM-Newton  telescope \citep{Gud2007}. These surveys identified $\sim$400 young stars and brown dwarfs over a sky coverage of more than 100~deg$^{2}$ in Taurus region \citep{Luh2010,Reb2010}.

\begin{figure*}
\centering
\includegraphics[angle=0,width=2\columnwidth]{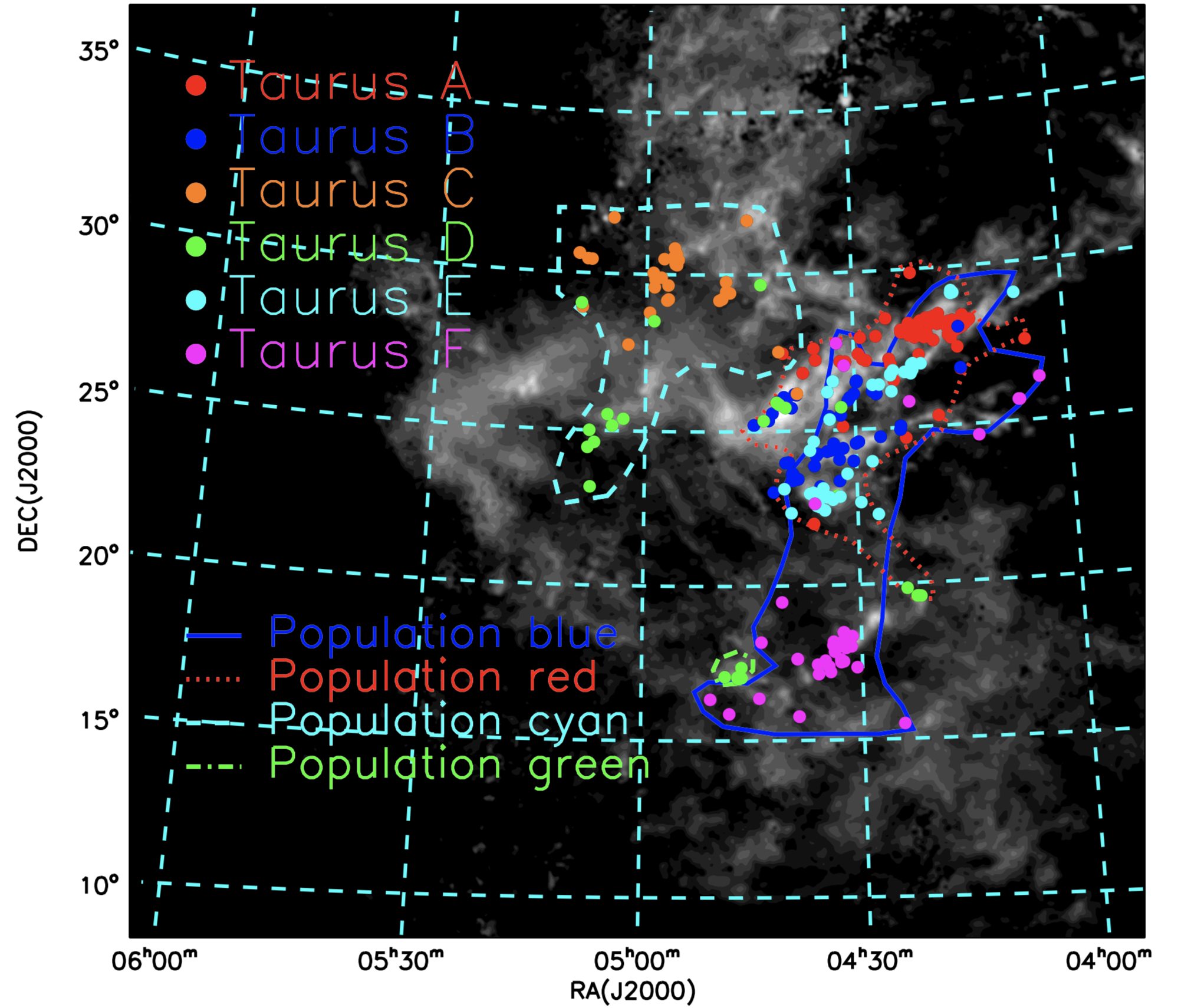}
\caption{The Taurus region and the previously identified YSO groups. Solid dots with different colors denote the distributions of the YSO groups of \cite{Roc2020}, while the lines coded with different colors are the locations of the YSO groups defined by \cite{Luh2018}. Background image are
the cumulative extinction map of \cite{Gre2019} to 300~pc.}
\label{Taurus}
\end{figure*}

In Taurus, the known young stellar objects (YSOs) tend to group within active star forming regions with dense molecular clouds  \citep{Bri2002,Luh2004,Luh2009,Esp2017}. With the astrometric data of {\em Gaia} DR2, \cite{Luh2018} found that these young stars can be grouped into four populations (i.e. Populations red, blue, green and cyan, see Figure~\ref{Taurus}).
The latest catalogue for the members in Taurus is from \cite{Esp2019}, in which 519 sources are listed.  With the latest catalogue, \cite{Roc2020} 
also studied the grouping of young stars in Taurus, and conclude that the members can be well divided into 6 populations (Figure \ref{Taurus}) with different distances and kinematics, indicating a complex star formation in this region.
Besides the young populations in Taurus, it has been known that there are "old" populations with ages of $\sim$10~Myrs distributed in the field of Taurus \citep{Sle2006,Kra2017,Zhang2018}. \cite{Luh2018} examine the kinematics and parallaxes of these "old" populations and conclude that most of them are not associated with the young populations in Taurus.


The previous studies on the stellar groups in the Taurus field mostly focused on the known young members, and a systematic search for stellar groups of older ages have not been performed. These groups could be relevant or irrelevant to Taurus star-forming regions and the known YSOs, and could improve our understanding of star formation history in the field of Taurus (not necessarily associated with the Taurus star-forming regions).  In considering of this, we carry out an extensive search for the stellar groups in the field of Taurus using the {\it Gaia} DR2 astrometric data. We organized the work as follows: 
we will describe data  in $\S$ 2, and delineate the data analysis in $\S$ 3. 
We will present the results in  $\S$ 4, followed by a discussion  
in $\S$ 5, and a summary of the work in $\S$ 6.





\section{Data} \label{sect2}
 
\begin{figure}
\centering
\includegraphics[angle=0,width=0.8\columnwidth]{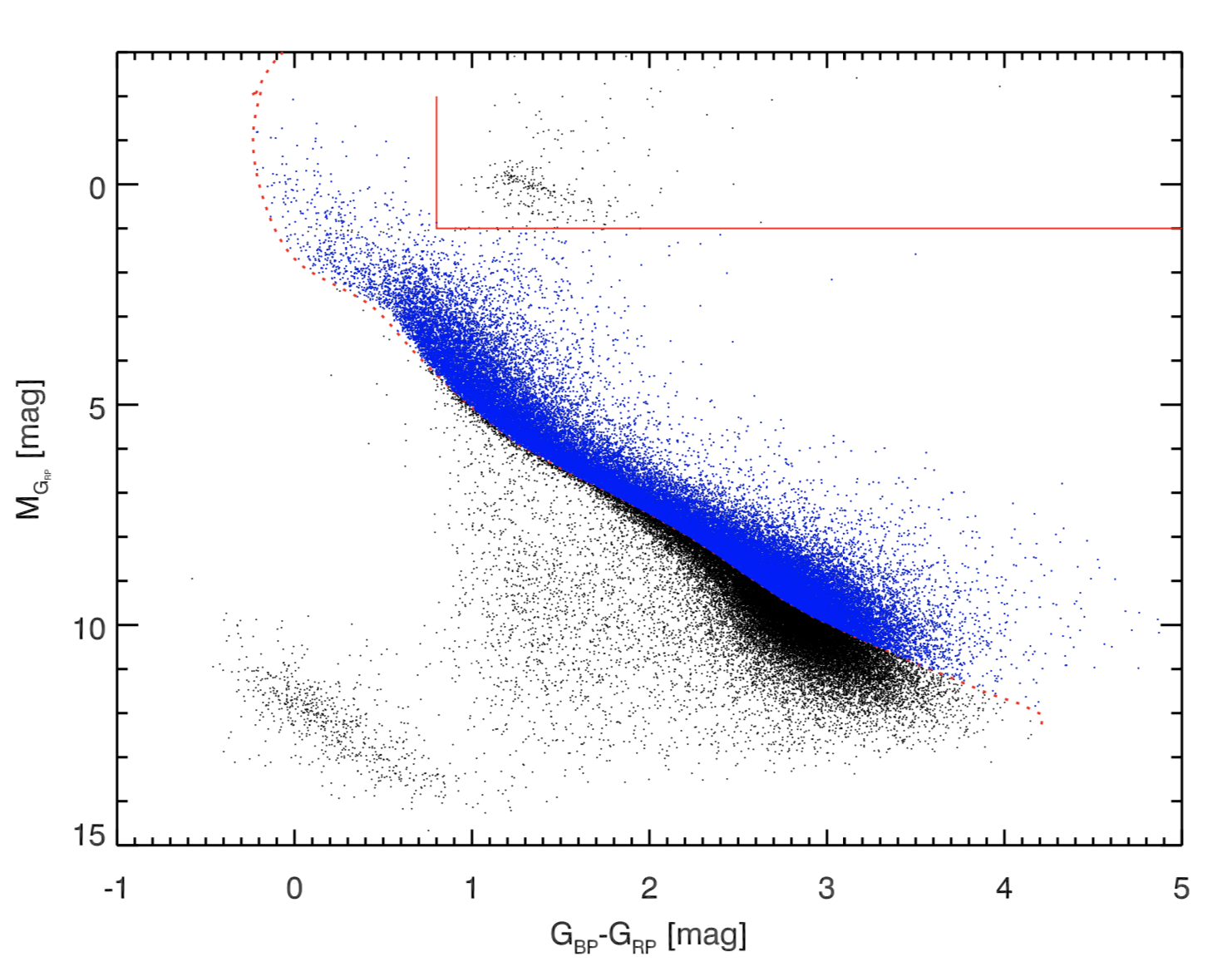}
\caption{The 100~Myr cut to eliminate the field main sequence stars. 
The red dotted line indicate the 100~Myr isochrone of PARSEC. The red solid line indicates the selection region, and blue dots denote stars selected in this 100~Myr cut.}
\label{cut}
\end{figure}

In order to fully explore the Taurus region, we  use a large searching area: $\rm 55^{\circ}\le RA \le 90^{\circ}$ and $\rm 10^{\circ}\le DEC \le 35^{\circ}$. Within this area, we extract the stars with  signal-to-noise ratios in parallax larger than 5 ( $\varpi/\sigma_{\varpi} \ge 5$). 
We only select the sources with the parallaxes between 3.33 and 10~mas, corresponding to 100-300~pc, in order to search for the groups with distance near the Taurus star forming region. In the northwest of Taurus, there is another star forming region, Perseus. We remove this region from the study in this work by excluding the sources  within the area of $\rm 55^{\circ}\le RA \le 70^{\circ}$ and $\rm 30^{\circ}\le DEC \le 35^{\circ}$. In our studied area, there is also a  known open cluster, Melotte~22. This cluster is easily identified in the space of proper motions (centered at $\mu_{\alpha}\sim$20 mas/yr and $\mu_{\delta}\sim$-45 mas/yr, \cite{Lod2019}). In this work, we limit $\mu_{\delta}$ between $-$40  and $-$10 \,mas/yr in order to (1) search for the potentially co-moving groups with Taurus and  (2) exclude the  Melotte~22 cluster from our sample. We are interested in searching for the young groups. To achieve this, we employ a color-magnitude diagram using the photometry from {\it Gaia} DR2 (see Figure \ref{cut}). We require that all the sources used in this work must be above the 100~Myr isochrone from \cite{Bre2012}. As noted in Figure \ref{cut}, there are a group of post-main sequence stars at the top-right side of the color-magnitude diagram. In order to remove them, we exclude the sources with $M_{G_{RP}}<1$ and $G_{\rm BP}-G_{\rm RP}>0.8$. The color-magnitude diagram shown in Figure \ref{cut} is not dereddened, then our selected sample could include some highly reddened old stars.  Thus, it is important to evaluate the reddening of individual sources which will be done later for the interesting objects, see \S\ref{Sect:extinction}.

\subsection{Spectra data}

The spectroscopic data used in this work are taken from LAMOST DR5.
LAMOST (The Large Sky Area Multi-Object Fibre Spectroscopic Telescope) 
is a 4m Schmidt telescope of the National Astronomical Observatories of China (NAOC), located at Xinglong Observing Station, China. With 4000 fibers on board the focus,  LAMOST can observe nearly 4000 spectra  simultaneously  in optical bands of $\sim 3900-9000$~\Angstrom,  at a resolution of $\sim 1800$ \citep{Cui2012}. In this work, we use the LAMOST DR5 dateset.


\subsection{Photometric data}

In order to construct the spectral energy distribution of each source and estimate its extinction, we used optical photometry in the $g$, $r$, $i$, $z$, and $y$ bands from  Pan-STARRS \citep{Cha2016} and $G$, $G_{BP}$, and $G_{RP}$ bands from Gaia DR2 \citep{GAIA2016,GAIA2017,GAIA2018}, near-infrared photometry in the $J$, $H$, and $K_S$ bands from the Two-Micron All Sky Survey \citep[2MASS, ][]{Skr2006}, near- and mid-infrared photometry in the $W1$ (3.4~$\mum$), $W2$ (4.6~$\mum$), $W3$ (12~$\mum$) and $W4$ (22~$\mum$) bands from  WISE  \citep[The Wide-field Infrared Survey Explorer]{Wri2010} allsky survey.



\section{Group searching and analysis} \label{sect3}

\subsection{Method: DBSCAN}

\begin{figure*}
\centering
\includegraphics[height=20cm]{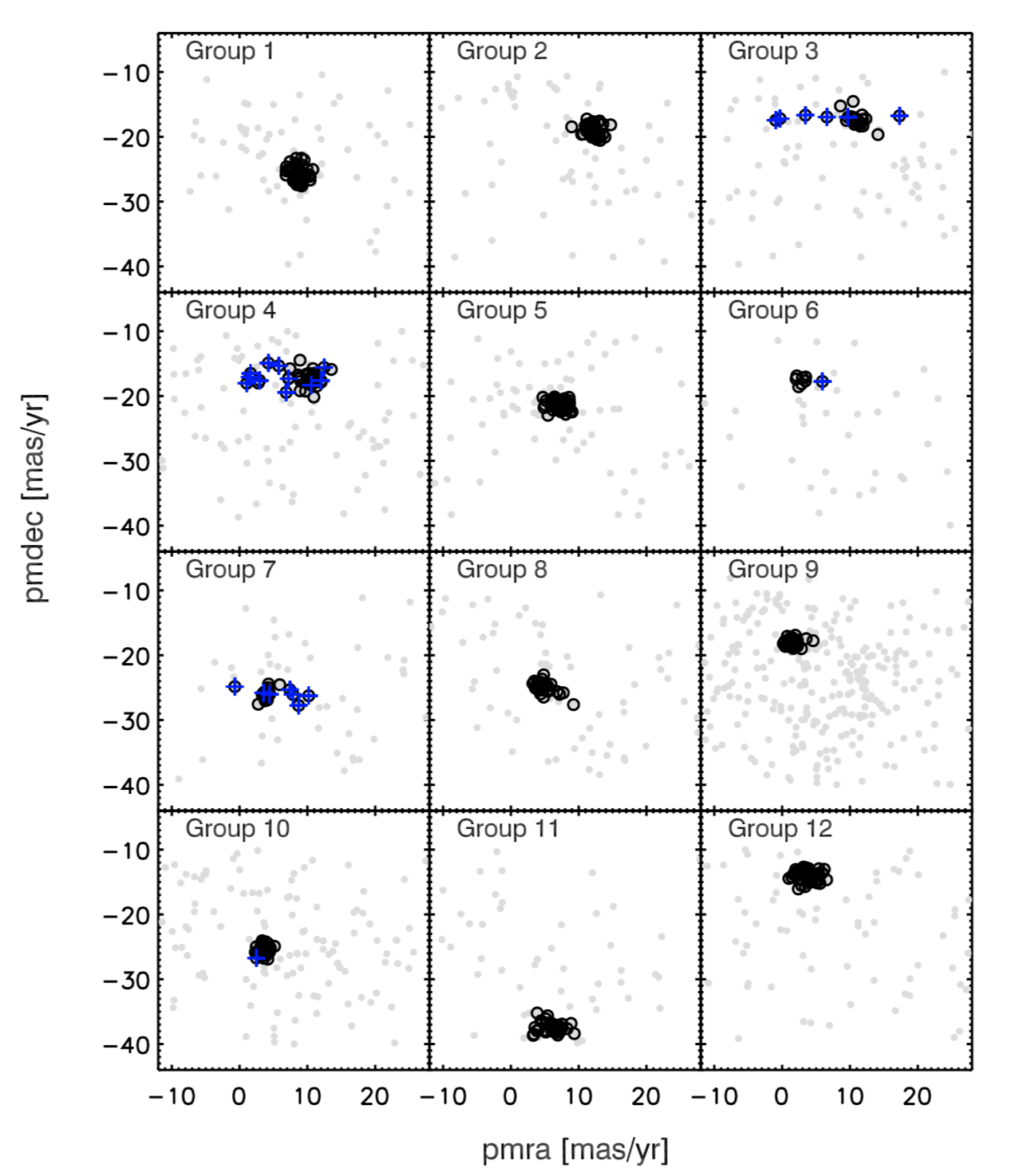}
\caption{The clustering of the identified groups 1--12 in the proper motion spaces. Gray solid dots are the surrounding stars within 15 pc from the center of each individual group, while black open circles mark group members located by DBSCAN. Note that in each panel we only include one group, and the members of other groups have been removed. The blue crosses mark the nonmembers that ruled out by CMD (see section \ref{sect:age}).}
\label{pm}
\end{figure*}

\begin{figure*}
\centering
\includegraphics[height=20cm]{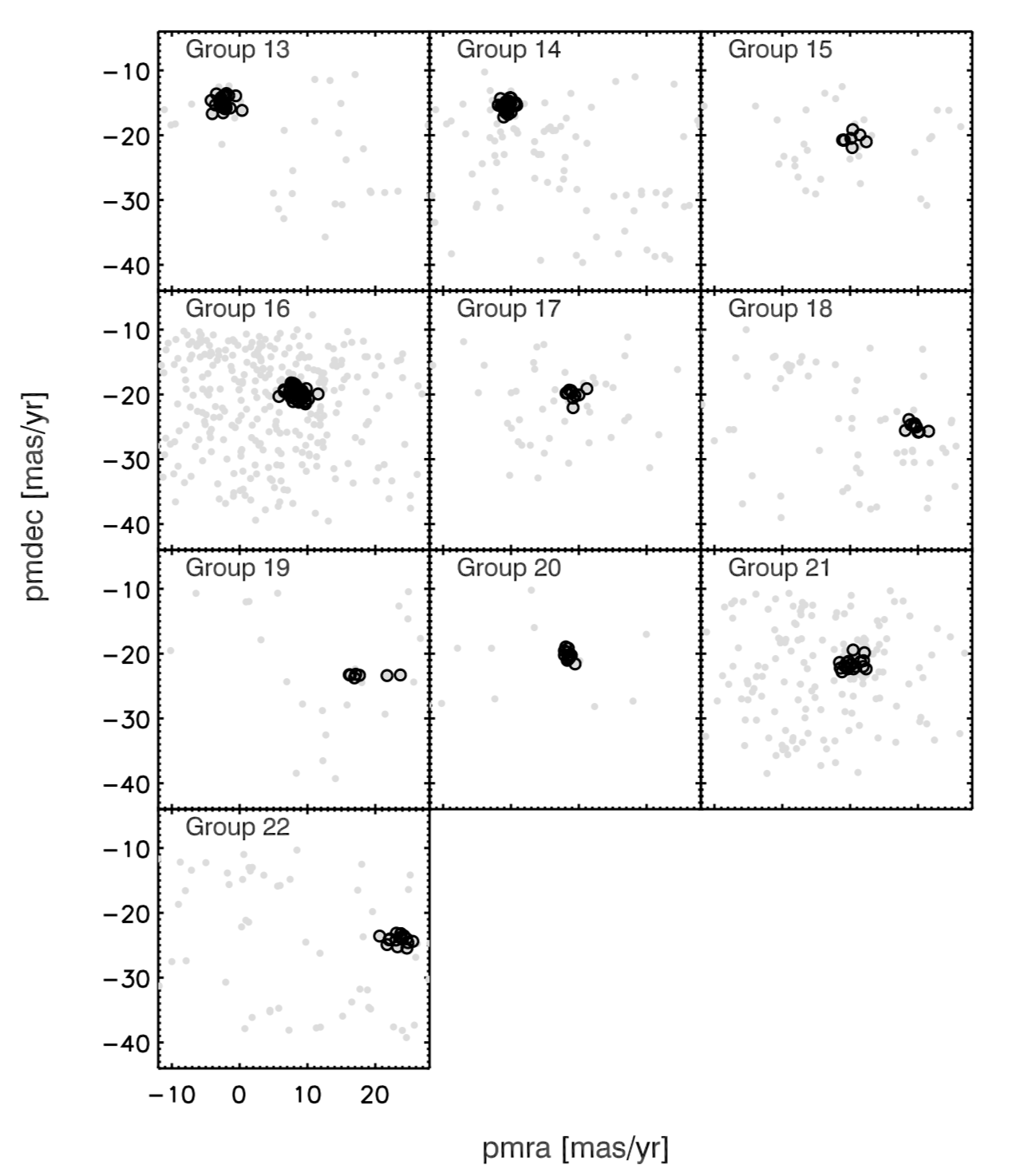}
\caption{Same as figure \ref{pm} but for Groups 13--22.}
\label{pm2}
\end{figure*}

In this work, We use DBSCAN to search groups in the Taurus field.  DBSCAN is a density-based clustering method. Its principal idea is if a point is belonging to a group, then it should be surrounded by members of the same group in the multi-dimensional space. The idea is realized by finding neighborhoods of 
data points exceed a given density threshold, which is defined as the minimum number of neighbors or data points ($minPts$) within a given search radius ($\epsilon$). The algorithm can be summarized as:

\begin{itemize}

\item[1.] With the given threshold ($minPts$ and $\epsilon$), starting from a random data point, the code will find all the points inside the radius $\epsilon$ (e.g. neighbourhoods of the data point). If the number of data points is greater than $minPts$, then all these data points will be regarded as a part of a ``cluster''.

\item[2.] From each of these data points identified  in Step 1, the code will repeat the step 1 to search for new data points upon the defined threshold above until no more new data points can be found. All the data points found in this step are also regarded as members of the cluster revealed in Step 1.

\end{itemize}

For DBSCAN, large $minPts$ values and small $\epsilon$ values are sensitive to those highly concentrated parts of the stellar groups, and on the contrary small $minPts$ and  large $\epsilon$ values  will include too much contamination from field stars, and mis-recognize groups. In this work, we perform the DBSCAN algorithm with the python package Sklearn of \cite{Ped2012} to a 5 dimensional normalized astrometric space that consists of $x$, $y$, $z$, $\mu'_{\alpha}$ and $\mu'_{\delta}$, where $x$, $y$, and $z$ are in cartesian space, $\mu'_{\alpha}$ and $\mu'_{\delta}$ refer to the tangential velocities in the right ascension and declination direction \footnote{For each parameter p of the 5 dimensional space, the normalization is done as: $p_{n}=(p_{i}-p_{min})/(p_{max}-p_{min}$), where $p_{max}$ and $p_{min}$ are the maximum and minimum value of $i$-th parameter $p_{i}$. }. We vary $minPts$ from 5 to 15, and $\epsilon$ from 0.01 to 0.1 to look for the reasonable values which can re-discover the YSO groups identified in Taurus. \cite{Roc2020} perform a group search among the spectroscopically confirmed members in Taurus, and find six populations, Groups A--F, with well defined parallax and proper motions. We find that by setting $minPts$ = 9 and $\epsilon=0.035$, the DBSCAN algorithm can re-discover these groups. With $minPts$ = 9 and $\epsilon=0.035$,  the DBSCAN algorithm find 22 groups in the whole region. However, we must stress that our group searching could ignore smaller and more sparse groups than the YSO groups since we have set the  $minPts$ and $\epsilon$ to be efficient for searching the YSO groups.




In order to minimize the influence of the projection effect and to remove the contamination in the group member from the field stars, we further refine the result in the above using a more rigorous criteria via two steps:

\begin{itemize}
\item[1.] For each group, we apply the DBSCAN algorithm to the stars within 15~pc from the center of the group by setting $minPts$=9 and varying  $\epsilon$ between 0.1 and 0.2. The normalization scale in this step is much smaller than the above one, and the $\epsilon$ used here is corresponding to 0.015 to 0.03 if using the same normalization scale as the above. In this step, we relatively reduce the search radius (smaller $\epsilon$) to mitigate the contamination in the group members from field stars. However, we could lose some members which are far from the group center. 

\item[2.] For each group, we only include the sources with proper motions within 
2$\sigma$ from the mean proper motions of the group (see Figure \ref{pm} and \ref{pm2}), where $\sigma$ is the standard deviation of the proper motions for the member candidates of the group.
\end{itemize}

\begin{figure*}
\centering
\includegraphics[angle=0,width=2\columnwidth]{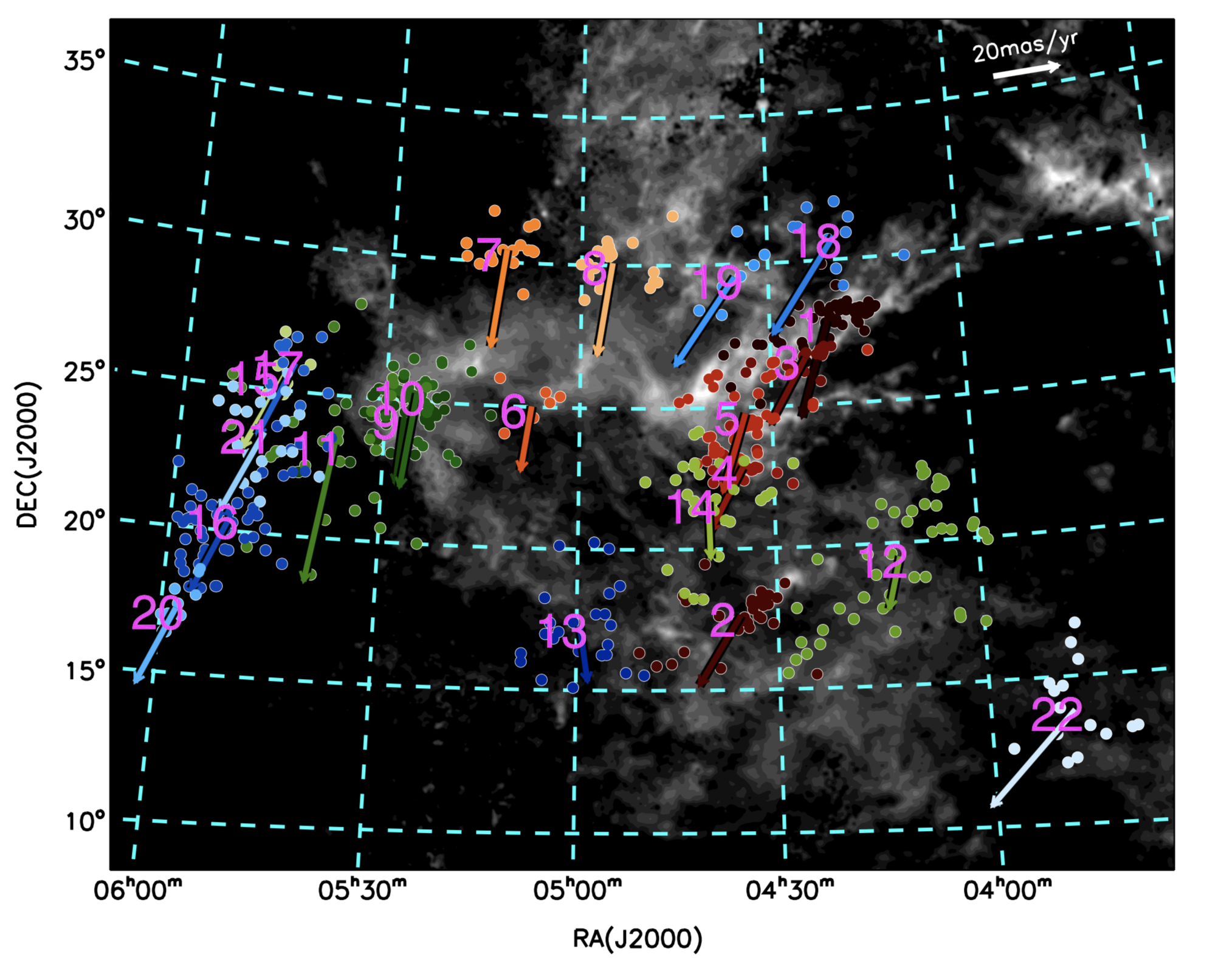}
\caption{Stellar groups identified in this work over plotted on the cumulative extinction map of \cite{Gre2019} to 300~pc. Each group is shown as different colors. The average proper motion for each group are show as arrows in the figure. }
\label{location}
\end{figure*}

During the refinements, we  have excluded about one hundred sources from our preliminary sample. Finally, we have 630 sources in our refined sample that are grouped into 22 populations (see Figure \ref{location} for their distribution in the sky). The standard deviation of these groups in the tangential velocity space is $\sim0.73$~km$^{-1}$. As a comparison,  We derive the the median value ($\sim2.33$~km$^{-1}$) of the standard deviations of tangential velocities derived from the stellar associations in \citealt{Gag2018} using Gaia DR2 data. This indicates that  our group method is very conservative and should have lost some group members. We will further refine their members of individual groups based on their locations in the dereddened M$_{G_{RP}}$ vs $G_{\rm BP}-G_{\rm RP}$ diagrams (see~Section~\ref{sect:age}).

We must stress that the group result could be influenced by the projection effect. It would be more reliable for searching stellar groups if using  the x, y, z, U, V, W spaces, where U, V, W velocities are the velocities in the x, y, z directions respectively. However, it  will severely cut down the number of stars as only bright stars ($G\ge13.0$) are released with radial velocities in {\em Gaia} DR2.

\subsection{spectral types}

\begin{figure}
\centering
\includegraphics[angle=0,width=1\columnwidth]{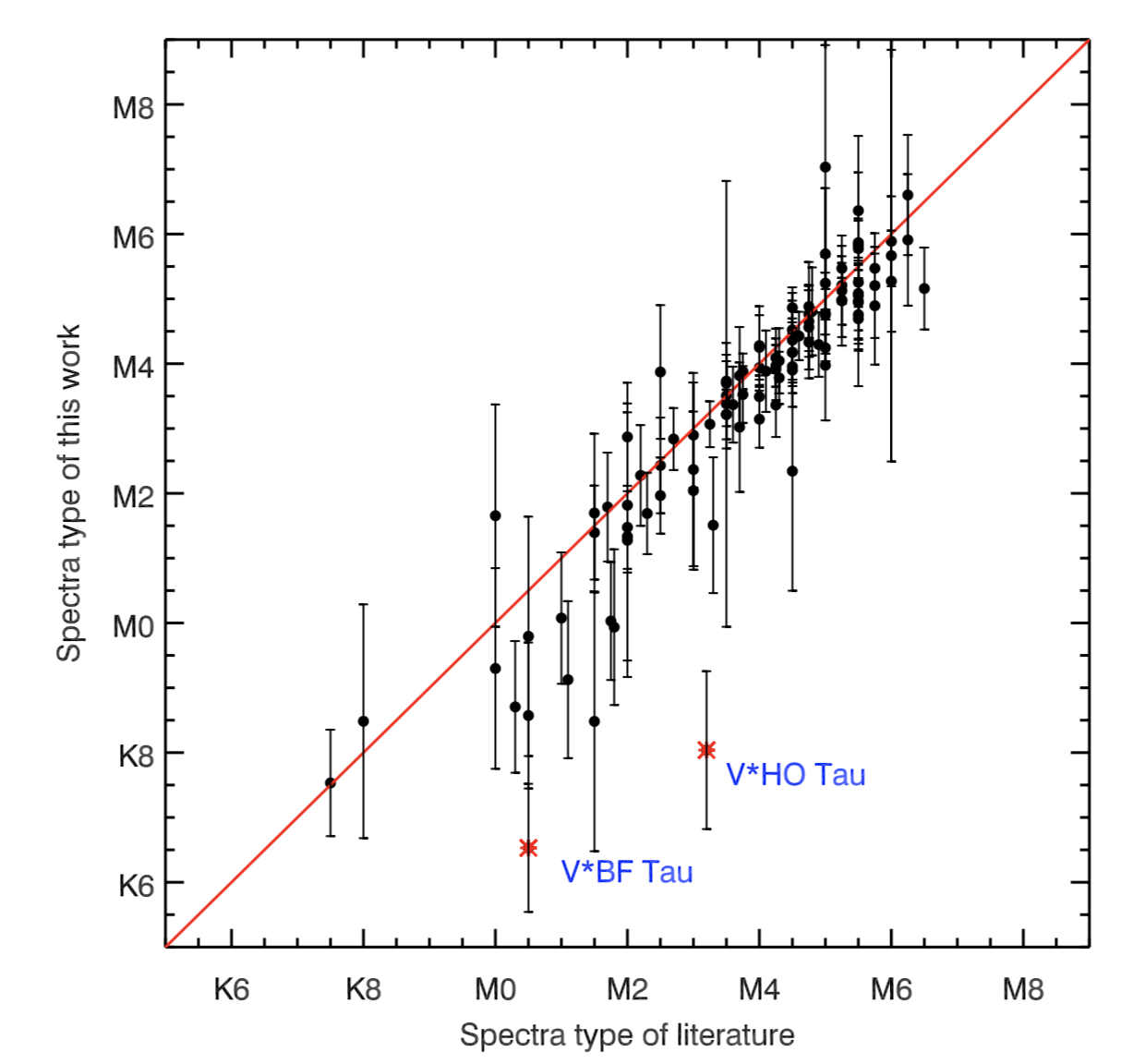}
\caption{The comparison of the spectral types between the literature \citep{Wic1996,Sle2006,Kra2017,Luh2018} and
those  derived from LAMOST spectra in this work. Red asterisks denote the two variable stars HO Tau and BF Tau.}
\label{spt_compare}
\end{figure}

We match the member candidates of each group with the known young stars with spectral types in the literature with a 2$''$ matching radius, and find 298 ones \citep{Sle2006,Kra2017,Esp2019,Can1993,Nes1995,Bia2012,Her2014,Gag2018b,Ker2019}. We also search for the spectra of our targets in LAMOST DR5, and retrieve the spectral data for 267 ones. Among them, 48 have been classified with the LAMOST pipeline, and the results are used in this work. For the other 219 sources without the spectral types from the LAMOST DR5, we classify their spectra with the  method described in \cite{Fan2017}. To evaluate the reliability of our spectral classification, we compared the spectral types of the 106 common sources with spectral type derived in this work and in the literature \citep{Wic1996,Sle2006,Kra2017,Luh2018}, see Figure~\ref{spt_compare}. The comparison shows there is no systematic difference between the spectral types  in the literature and in this work. We note that two sources, BP Tau and HO Tau, show more than 2$\sigma$ difference on the spectral types. The large difference on the spectral types for an accreting young star is very common, and can be due to variable optical veiling on the spectra \citep{2020arXiv200911995F}, since the spectral classification usually does not consider veiling effect.

In total, we have  spectral types for 427 sources in our sample, and 298 ones from literature and 129 derived from the LAMOST spectra.

\subsection{Extinction}\label{Sect:extinction}

\begin{figure*}
\centering
\includegraphics[angle=0,width=2\columnwidth]{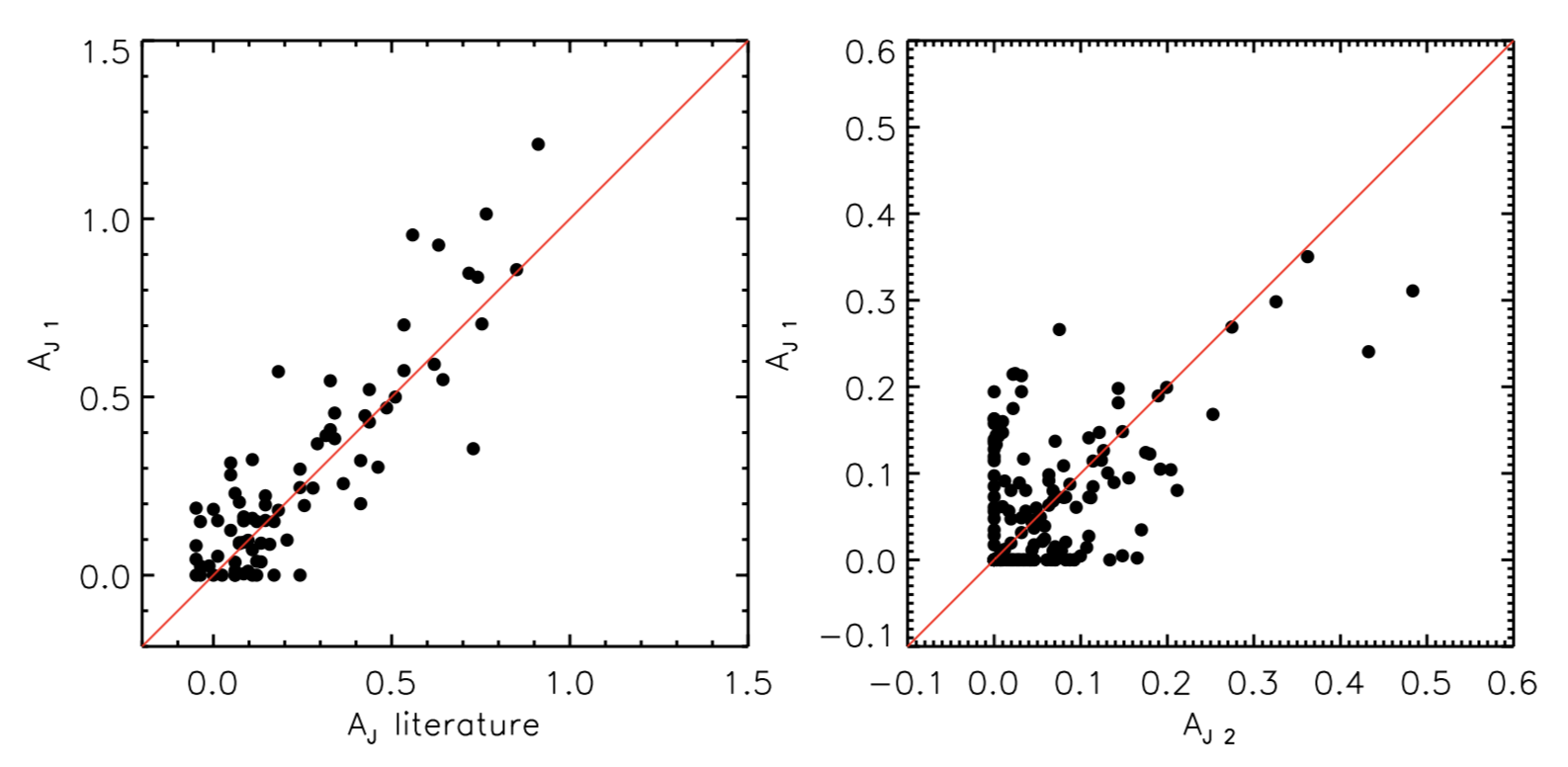}
\caption{The extinction comparison. The left panel shows the comparison between the extinctions derived from the method we applied to stars with spectral types ($A_{J 1}$) and the literature \citep{Her2014}. Right panel is the comparison of the extinctions derived from the two methods of this work ($A_{J 1}$ derived from the method we applied to stars with spectral types, while $A_{J 2}$ for method of no spectral types). The target stars are members of comparatively older groups (age $\ge$5\,Myr) that with known spectral types. The red solid line shows the 1:1 relation.}
\label{AJC}
\end{figure*}

Our sample contains about three hundreds well studied young stars in Taurus, and their published extinction are adopted in this work \citep{Her2014,Esp2019}. For those without estimates of extinction in the literature, we derive their extinctions. For the ones with spectral types, we convert their spectral types to effective temperatures ($T_{\rm eff}$) using the conversions in \cite{Fan2017}, which are from \cite{Pec2013} for stars earlier than M4 and from \cite{Her2014} for stars later than M4 type. We then achieve their intrinsic colors for $G_{RP}-J$ corresponding to their $T_{\rm eff}$ by interpolating the model colors for the 10\,Myr isochrone from PARSEC \citep{Bre2012}, and derive the extinction of individual sources using the average extinction law  with the total-to-selective extinction ratio $R_{V}=3.1$  from \cite{Wan2019}. We verify this method by comparing our results with those in \cite{Her2014} for the common sources. The comparison shows that both agree well with each other, see the left panel in Figure~\ref{AJC}.

For the sources without spectral types, we perform  a least-square fit to the observed colors, taking the extinction and $T_{\rm eff}$ as free parameters. The colors used in the fitting are a combination of the broad-band photometry in {\em Gaia}  $G_{BP}$, $G_{RP}$ bands and 2MASS  $J, H, K_{s}$ bands. For each  $T_{\rm eff}$ we obtain  the model colors for the 10\,Myr isochrone from PARSEC \citep{Bre2012},  redden them using the same extinction law with $R_{V}=3.1$  from \cite{Wan2019}, and then compare the reddened colors to the observed ones. The best-fit to the observed colors yields the extinction used in this work. The extinctions derived in this method are mostly for the diskless sources. To verify this method, we collect a sample of the sources with spectral types in the old groups and estimate their extinctions using the above two methods. We compare the extinctions from the two methods in the right panel in Figure~\ref{AJC}, which shows a good agreement between them.

We further verify our derived extinction by repeating the above procedure by using  the model  colors for the 100\,Myr isochrone from PARSEC \citep{Bre2012}. The derived extinctions are consistent with the above ones using the model  colors for the 10\,Myr isochrone from PARSEC. The $J$-band extinction of each source used in this work is listed in Table~\ref{Tab:star_inform}.

\begin{figure*}
\centering
\includegraphics[angle=0,width=2\columnwidth]{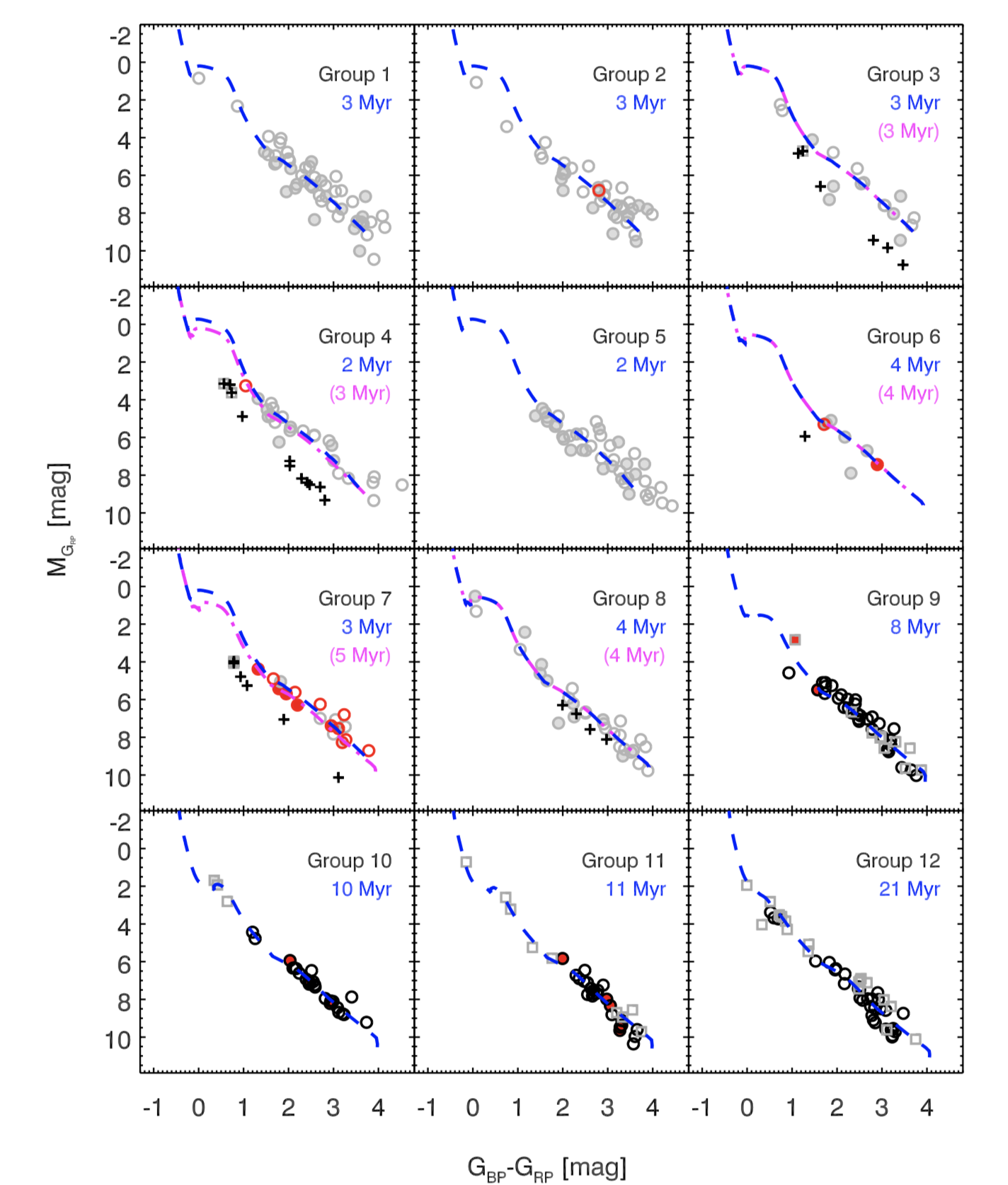}
\caption{Best-fit isochrones (blue dashed lines) for the stellar groups 1--12 in dereddened M$_{G_{RP}}$ vs $G_{\rm BP}-G_{\rm RP}$ diagrams. The purple dashed-dotted lines in the panels of group 3, 4, 6, 7 and 8 denote the isochrone fit result without removing the contamination of field stars. New identified YSOs and disk bearing stars of this work are denote as red open circles and red filled circles respectively. Gray open circles are YSOs catalogued by \cite{Esp2019}, while the gray rectangles show the known intermediate-age PMS stars in 
the literature (e.g. \citealt{Sle2006} and \citealt{Kra2017}). The disk-bearing stars confirmed by \cite{Esp2019} are denote as gray filled circles. Black crosses mark the likely contaminants of field stars.}
\label{CMD1}
\end{figure*}

\begin{figure*}
\centering
\includegraphics[angle=0,width=2\columnwidth]{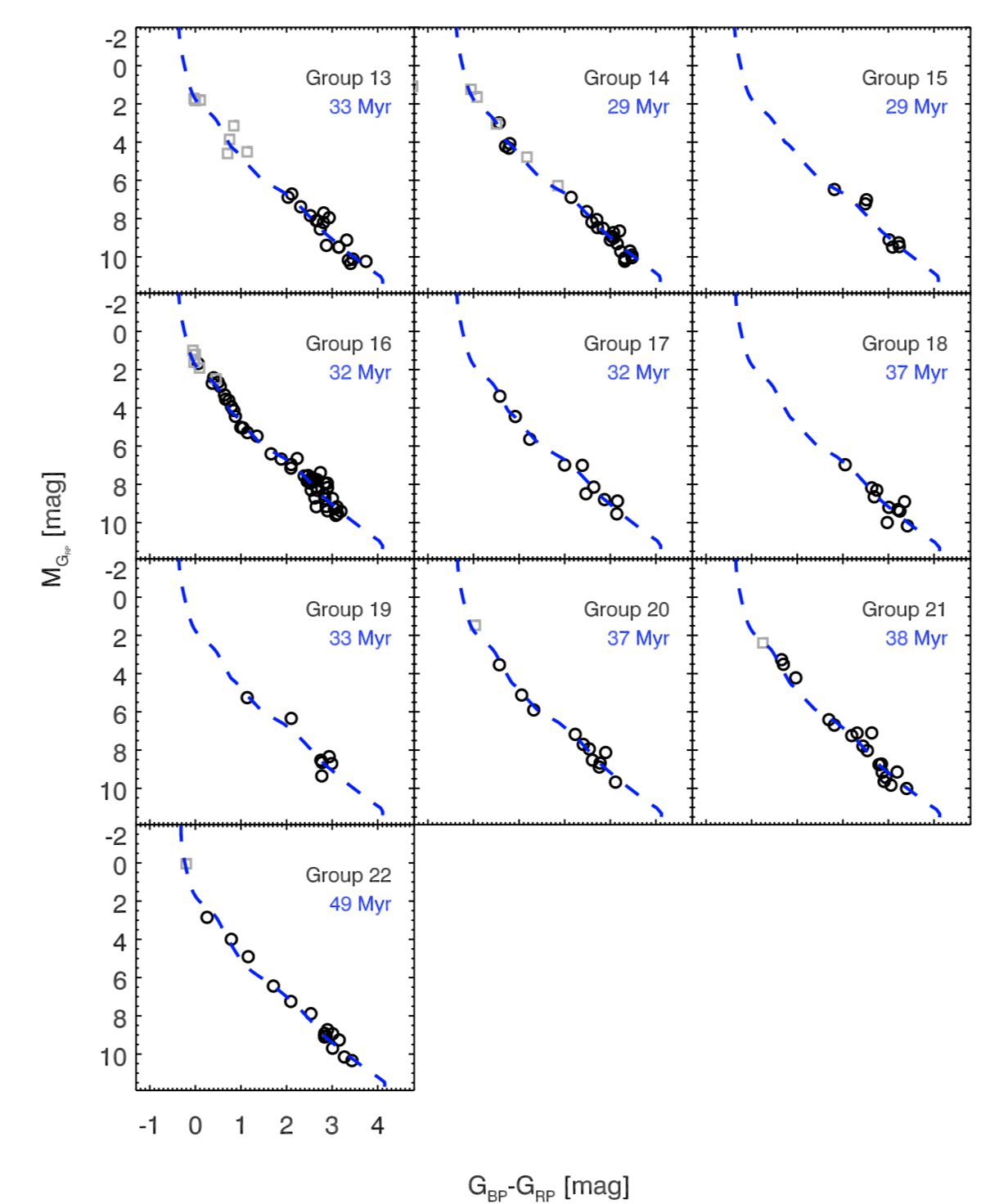}
\caption{Same as figure \ref{CMD1}, but for Groups 13-22.}
\label{CMD2}
\end{figure*}

\section{result}

\subsection{The ages of groups}\label{sect:age}

In Figures~\ref{CMD1} and~\ref{CMD2}, we show the  dereddened M$_{G_{RP}}$ vs $G_{\rm BP}-G_{\rm RP}$ color-magnitude diagrams (CMD) of individual groups identified in Section~\ref{sect3}. We fit the dereddened M$_{G_{RP}}$ vs $G_{\rm BP}-G_{\rm RP}$ diagrams  with the model isochrones of the PAdova and TRieste Stellar Evolution Code (PARSEC, \citealt{Bre2012})\footnote{Notice that, in the isochrone fitting process, only 
stars with the flux errors in $G_{BP}$ and $G_{RP}$ bands less than 10\% are included.}. The best-fit isochrone for each group is achieved by minimizing the mean distance of its member candidates to the isochrones. The age of the best-fit isochrone is adopted as the age of the group. From the fitting, 8 groups (Groups 1-8) have ages of 2--5\,Myr, while the other 14 groups have ages ranging from 8 to 49\,Myr.

\begin{figure}
\centering
\includegraphics[angle=0,width=1\columnwidth]{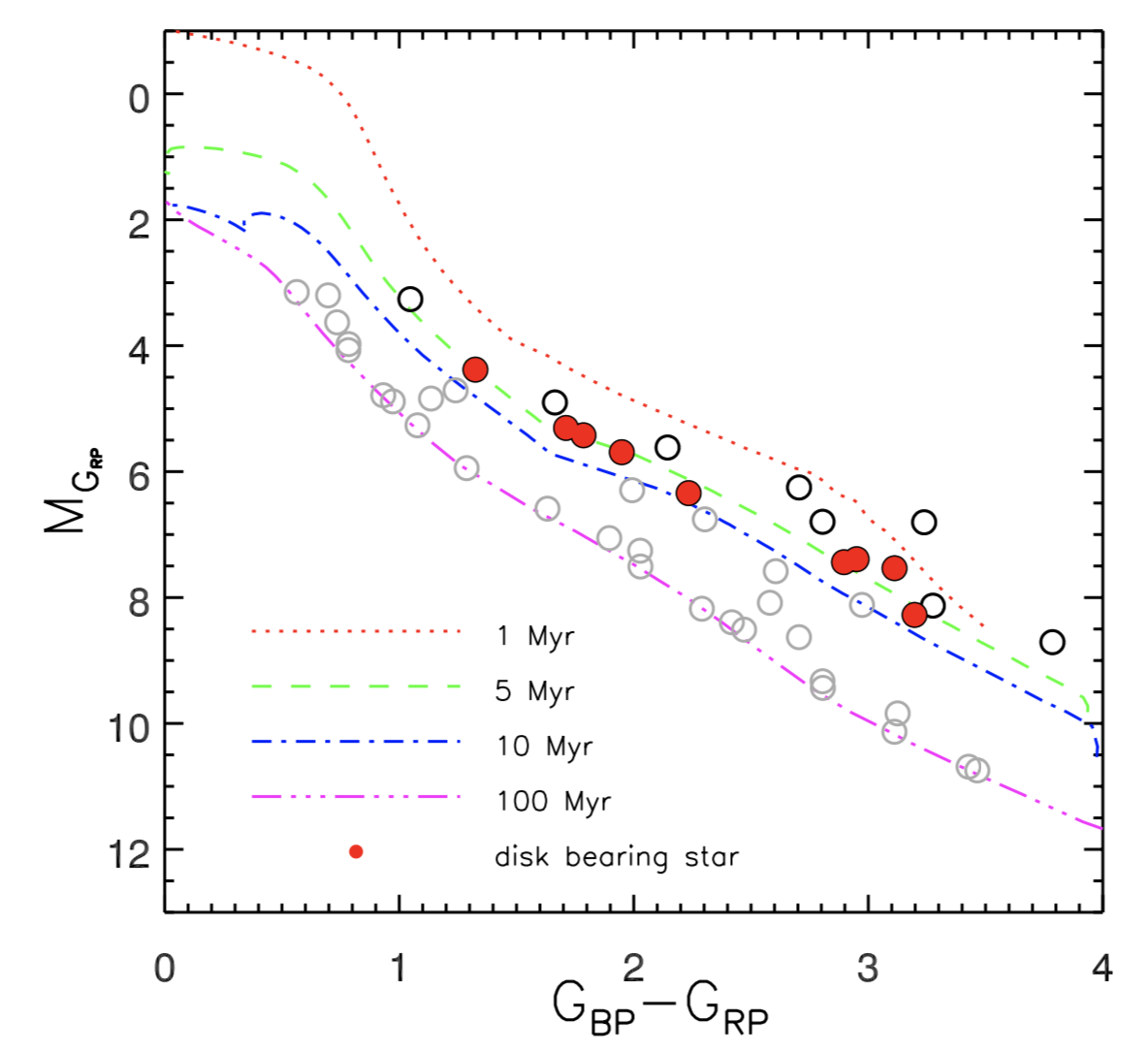}
\caption{The dereddened M$_{G_{RP}}$ vs $G_{\rm BP}-G_{\rm RP}$ diagrams for the newly discovered member candidates (black open circles for those without disks and red filled circles for those with disks) in young groups. The gray open circles show the sources with ages old than 10~Myr and are excluded as the members of young groups. The isochrones are taken from the PARSEC models.}
\label{verify}
\end{figure}

In 6 young groups, Group 2, 3, 4, 6, 7, and 8, we have found 47 new member candidates in total. We note that many of them  locate below the best-fit isochrone for each young group and could be contaminators from "older" populations in the field. This can be even clearer in Figure~\ref{verify} where we plot the 47 member candidates together in the  dereddened M$_{G_{RP}}$ vs $G_{\rm BP}-G_{\rm RP}$ diagram. In the figure, about 50\% of them locate near the 100~Myr old isochrone. In this work, we only inlcude the ones above 10~Myr old isochrone as the probable members of young groups. This criterion is defined as a compromise between reducing the contamination from the "older" populations and the large spreads in the  dereddened M$_{G_{RP}}$ vs $G_{\rm BP}-G_{\rm RP}$ diagrams for young groups. According to this criterion,  we include the 17 sources as the members of the young groups, 13 of which are in Group 7, and  
exclude  30 the other ones in the further analysis and discussion.

After excluding the contamination, we re-fit the dereddened M$_{G_{RP}}$ vs $G_{\rm BP}-G_{\rm RP}$ diagrams of the 5 young groups using the PARSEC isochrones. For Groups 2, 3, 6, 8, we derive the similar ages as before, and for Groups 4 and 7, we obtain the younger ages than before, 2\,Myr vs.3\,Myr, and 3\,Myr vs.5\,Myr. After excluding the contaminators, in Figures~\ref{CMD1} we can still see that some sources are located below the best-fit isochrones and look like old populations. All these sources  are known young stars in the literature and harbor disks (see Section~\ref{disk}). The locations of these sources in CMDs could be due to accretion activities, the disk orientation, etc. \citep{2020arXiv201114483F}. Compared with the young groups, the "old" groups show well defined loci in the CMDs, and their ages can be constrained very well with the PARSEC isochrones. The age of each group is listed in Table~\ref{Tab:Tab:aggt_info}.

We also fit the observed M$_{G_{RP}}$ vs $G_{\rm BP}-G_{\rm RP}$ diagram using the PARSEC isochrones with an assumption that all member candidates of each group have the same extinction\footnote{Noted the assumption on the extinction may be improper for the young groups where the extinction from circumstellar material and its parental cloud vary from star to star.}. As a comparison, the ages of the best-fit isochrones are also listed in Table~\ref{Tab:Tab:aggt_info}. In general, the ages estimated in two ways are consistent with each other.

\begin{figure}
\centering
\includegraphics[angle=0,width=1\columnwidth]{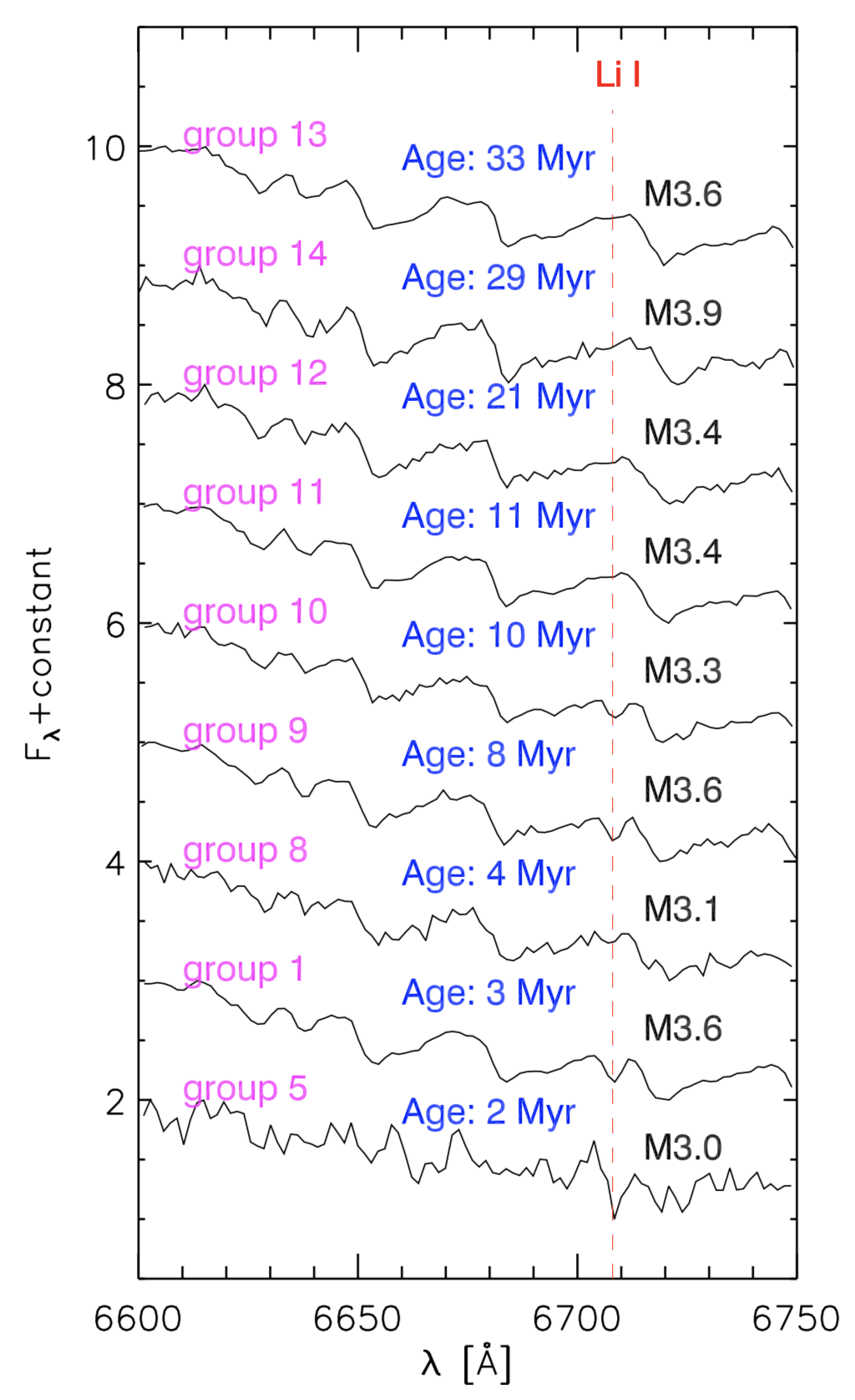}
\caption{LAMOST low-resolution spectra for $\sim$M3--M4 type stars at different ages. The red dashed line denotes the Li\,I 6708\,\AA\ line.}
\label{Li_age}
\end{figure}

We verify our age estimate using the strength of the Li\,I absorption line at 6708\,\AA\, which is a good indicator of stellar ages \citep{Soderblom2014}. In Figure \ref{Li_age}, we show the Li\,I absorption line in the LAMOST spectra of the sources with spectral types around M3. Although the spectral resolution is relative low (R$\sim$1800), we can clearly see the rapid Li depletion around 10-30\,Myr for the M3 type young stars, which is consistent with the results in the literature (see \citealt{Zuckerman2004} for a review).

\subsection{Circumstellar disks} \label{disk}

\begin{figure*}
\centering
\includegraphics[height=6cm]{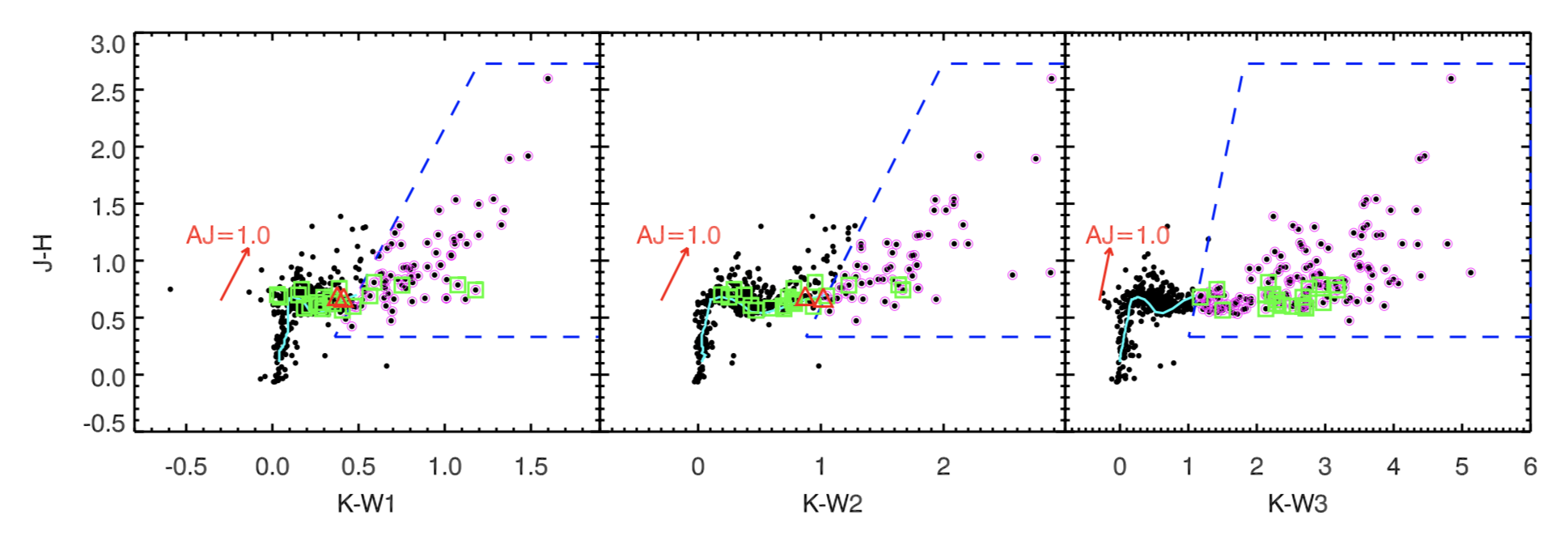}
\caption{Color-color diagrams for the members (black dots) of all the groups identified in this work. Cyan solid curve in each panel shows the intrinsic colors of PMS stars from \cite{Pec2013}, while red arrow denotes the extinction vector of $A_{J}=1.0~$mag. 
Purple open circles in each panel display the sources with infrared excess emission in the used WISE band in that panel, and the green boxes mark the 19 disk-bearing stars newly identified in this work. The two disk bearing stars which are not confirmed in this work due to the lack of W3 band photometries are marked as red rectangles.}
\label{CCD}
\end{figure*}

\begin{figure*}
\centering
\includegraphics[angle=0,width=2\columnwidth]{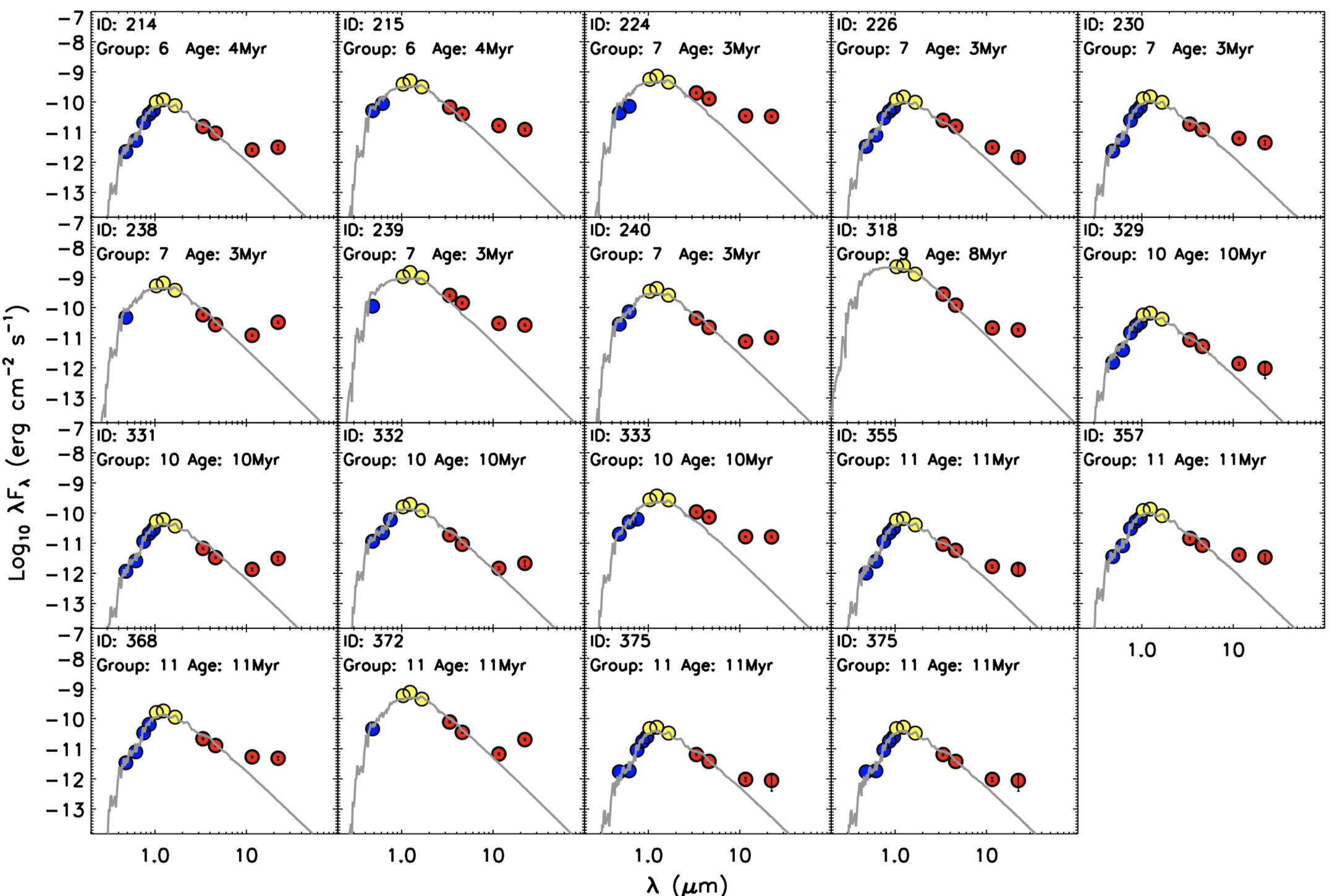}
\caption{The SEDs of the 19 newly discovered disks. Blue solid circles denote the photometries
of Pan-STARRS (The Panoramic Survey Telescope \& Rapid Response System,  \cite{Cha2016}), while yellow and red solid circles indicate 2MASS and ALLWISE bands respectively. In each panel, the gray line indicates the photospheric emission level.}
\label{SED}
\end{figure*}

In this work, we use the infrared photometry from {\em 2MASS} and AllWISE Catalog to search for the objects with disks. In both the catalogs, we extract photometry for 609 common stars. In Figure \ref{CCD}, we show their infrared color-color diagrams used to identify the sources with disks. In the figures, the stars to the bottom-right side of the reddening vector are too red to be explained by the reddening of diskless stars, suggesting there are substantial excess emission above the photospheric level in these wavelengths, which is most likely due to emission from the disks surrounding them. Thus, we identify these sources as stars with disks. In this way, we identify 123 reliable disk-bearing sources (see Figure \ref{CCD}), including 104 known ones confirmed in the literature \citep{Mah2002,Reb2011,Eis2004,Esp2014,Men2015,Esp2019}\footnote{Among the 104 sources, Sources 595 and 596 are two A-type stars belonging to one binary system (A2+A7) with a circumstellar disk around the A2-type star, Source 596 \citep{Dunkin1998}.}. We compare our disk classification with the result in \cite{Esp2019}, and notice that we have missed 20 disk-bearing stars. Among them, 14 ones only show the infrared excess emission in {\it WISE} W4 band or the Spitzer 24 $\mum$ band, and thus are not identified in this work\footnote{The photometry in Wise W4 band is not used to identify the disk-bearing star in this work.}. Among the other 6 sources, three of them (Sources 63, 250, and 255) can not be cross-matched with the sources in the ALLWISE catalog within 2$''$.  Another two stars (Sources 19 and 118, red rectangles in Figure \ref{CCD}) locate near the boundary in the left two panels in Figure \ref{CCD} which we use to identify the sources with infrared excess emission, and might have weak infrared excess emission in those {\it WISE} bands. Unfortunately the two sources have no photometry in {\it WISE} W3 band. For the reminding one source (Source 91 or V710 Tau A), it is one component of a binary system (V710~Tau), and the {\it WISE} data cannot resolve the system. In this work, we list the 20 sources as harboring disks in Table~\ref{Tab:star_inform}.  The other 19 disk-bearing sources are firstly revealed in this work. In Figure \ref{SED}, we show the spectral energy 
distributions (SED) of these sources. Among these sources, 8 are in young groups (Groups 6, 7 and 8). Interestingly, we also find 11 disks in the relatively older Groups (8 to 11\,Myr), 2 in Group 9, 4 in Group 10, and 5 in Group 11.


\subsection{Accretion}


Classical T-Tauri Stars (CTTS) and Weak-Line T-Tauri Stars (WTTS) can be distinguished from the H$\alpha$ emission \citep{Whi2003}. CTTSs usually show strong and broad  H$\alpha$ emission line due to the accretion process, and WTTSs present  weak and narrow H$\alpha$ emission line due to the chromospheric activity. In this work, we use the criteria from \cite{Fan2009} to divide the stars with LAMOST spectra into WTTSs and CTTSs based on their H$\alpha$ equivalent widths ($EWs$). Among the  224 stars with H$\alpha$ emission in their spectra, 38 are CTTSs. The results are listed in Table~\ref{Tab:star_inform} (column 11). Among the CTTSs, 37 sources are in the groups with ages younger than 5\,Myr, and one (Source~332) is in  Group 10 at an age of 10\,Myr (see the discussion in Section~\ref{Sect:Intermediate-age-Groups}).

\begin{figure}
\centering
\includegraphics[angle=0,width=1\columnwidth]{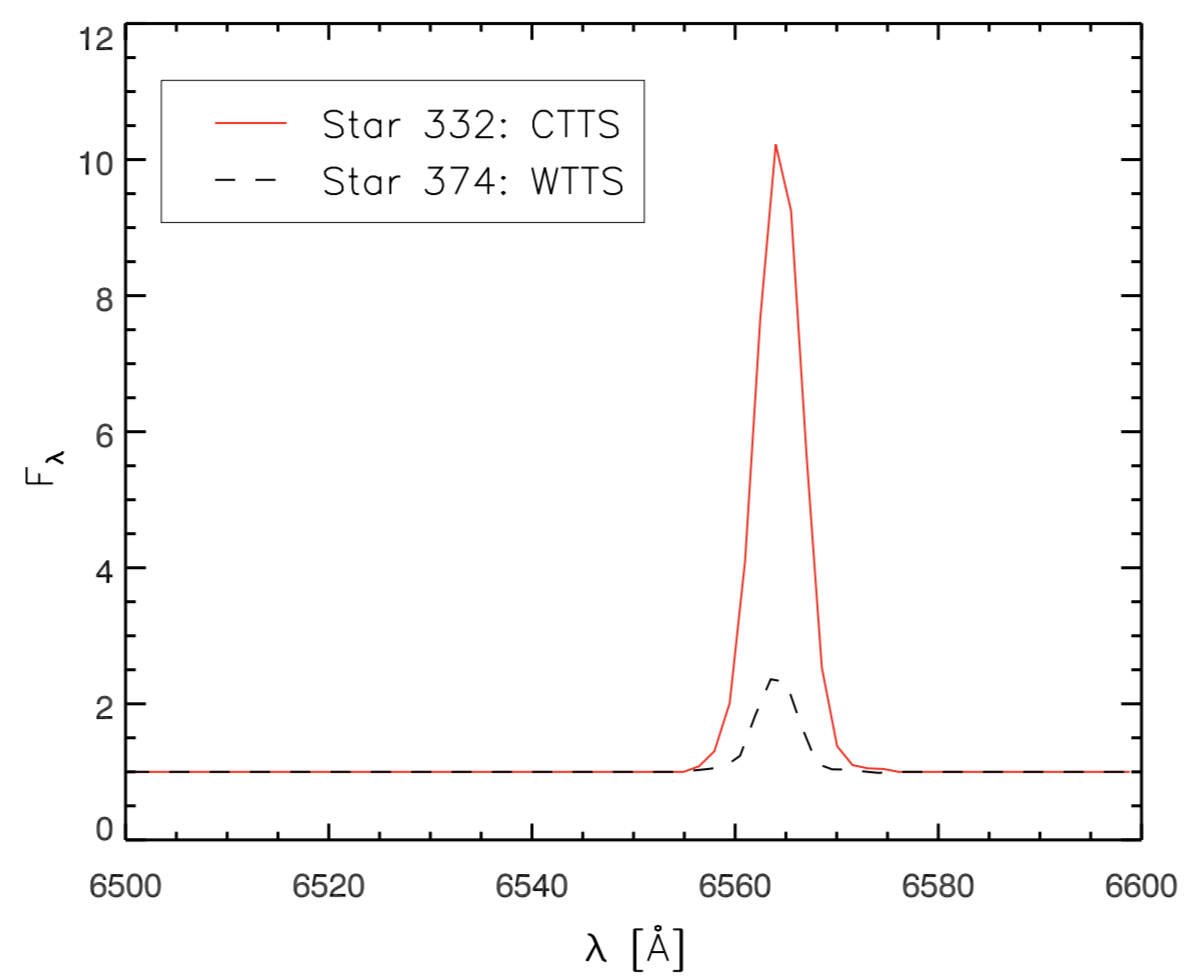}
\caption{The H${\alpha}$ emission comparison between Source 332 (red solid line) and a WTTS star (Source 374, black dashed line) of same spectral type.}
\label{332}
\end{figure}

In the relative older groups, we discover 11 sources with circumstellar disks, see Fig.~\ref{SED}. Among the 11 sources, we have LAMOST spectra for 4 of them, Sources 331 (Group 10), 332 (Group 10), 355 (Group 11) and 368 (group 11). Based on their H$\alpha$ $EWs$, Source 331, 355 and 368 are classified as  WTTSs and Source~332 is a CTTS. The H$\alpha$ $EW$ of Source~332 is about $-$53\,\AA, which is much stronger than the threshold ($-$18\,\AA) used to distinguish CTTSs from WTTSs. Figure \ref{332} compares its H$\alpha$ line with that of a WTTS star (Source 374) with a similar spectral type.
The H$\alpha$ line profile of Source~332 is well resolved in the LAMOST spectra.  We fit its H$\alpha$ line profile using a Gaussian function and derive the full width of half maximum ($FWHM$) of 243\,\kms. With an assumption that the intrinsic 
H$\alpha$ line profile is a Gaussian function and after deconvolved from the spectral resolution ($R\sim$1800), the intrinsic full width at 10\% maximum ($FWHM_{10\%}$) of the line profile is 343\,\kms, which gives an accretion rate of $\sim3\times10^{-10}$\,$M$\accunit using the relation between the $FWHM_{10\%}$ and accretion rates from \cite{Natta2004}. Source~332 shows a SED, typical for an evolved disk, with no infrared excess in WISE-W1 and W2 band, and weak excess emission in WISE-W3 and W4 bands. Thus, we might witness the accretion process at the latest stage of disk evolution.

\section{Discussion}\label{discuss}

The ages of our groups range from 2 to 49~Myr. According to their ages, we divide them into two categories: young Groups (2--4\,Myr) and old groups (8--49\,Myr). A discussion of these groups and their relevance to the results in the literature is as follows. 


\subsection{The Young Groups of this work}





\begin{figure}
\centering
\includegraphics[angle=0,width=1\columnwidth]{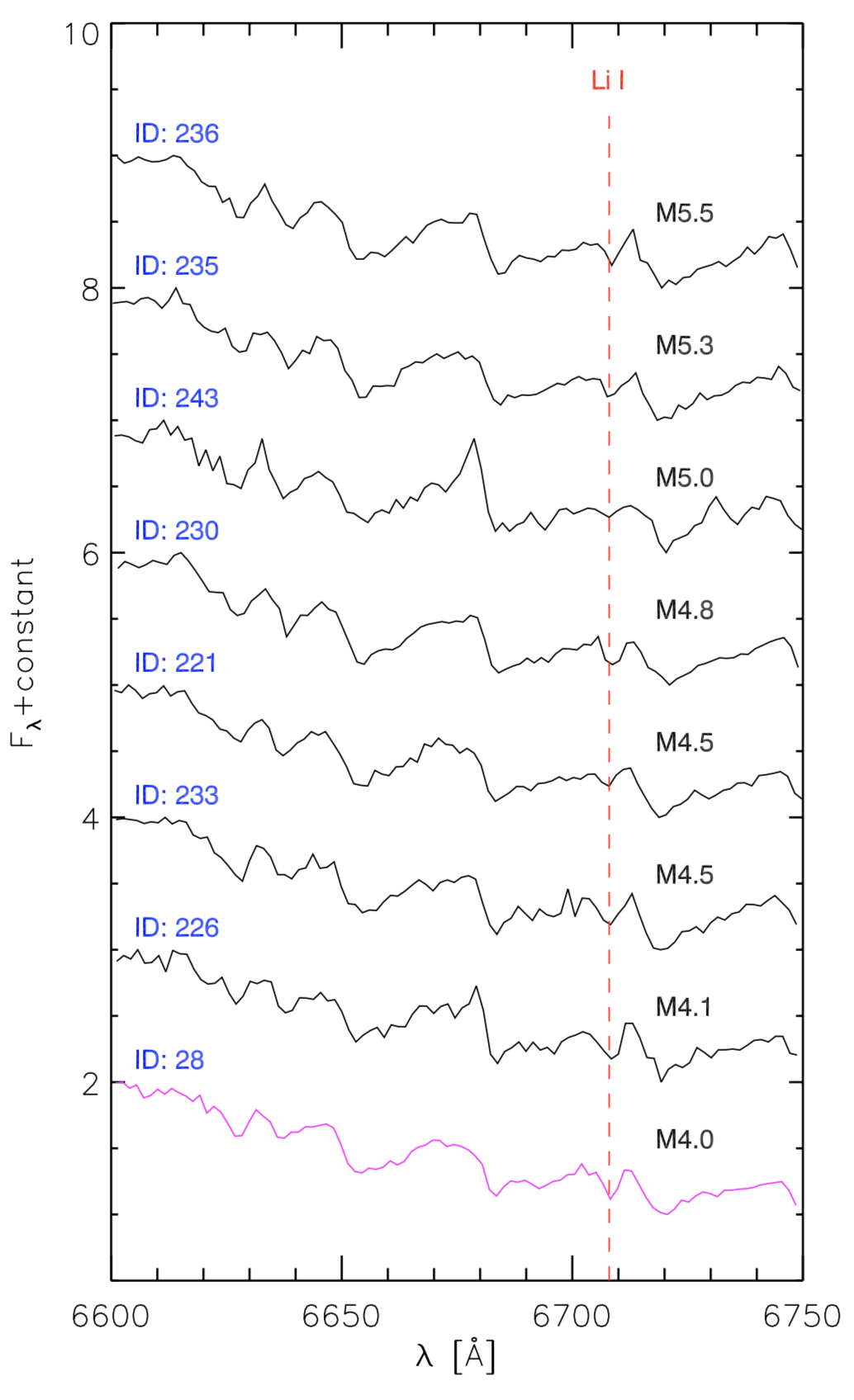}
\caption{LAMOST spectra for 7 members in Group 7 (black) and one in Group~1 (Source 28, purple). Red dashed line denotes the Li\,I absorption line at 6708\,\AA. }
\label{group7}
\end{figure}

In this work, we identify 8 groups, Group 1-8, with ages younger than $\sim$4\,Myr.  Among them, Group 7 have the largest members newly identified in this work. In this group, there are 19 sources and 13 are new. In Group~7, we have 7 M4--M5 type YSOs with LAMOST spectra. In Figure \ref{group7} we show the Li\,I absorption line at 6708\,\AA\ for 7 sources. As a comparison, we also show one M4.0 type young stars (Source 28) in Group~1 at the similar age (3\,Myr) to that of Group~7.  The strengths of the Li\,I absorption lines of the 7 members in Group~7  are comparable to the one of Source 28, see Figure \ref{group7}. This supports that Group~7 is at the similar age to that of Group~1.

\begin{figure}
\centering
\includegraphics[angle=0,width=1\columnwidth]{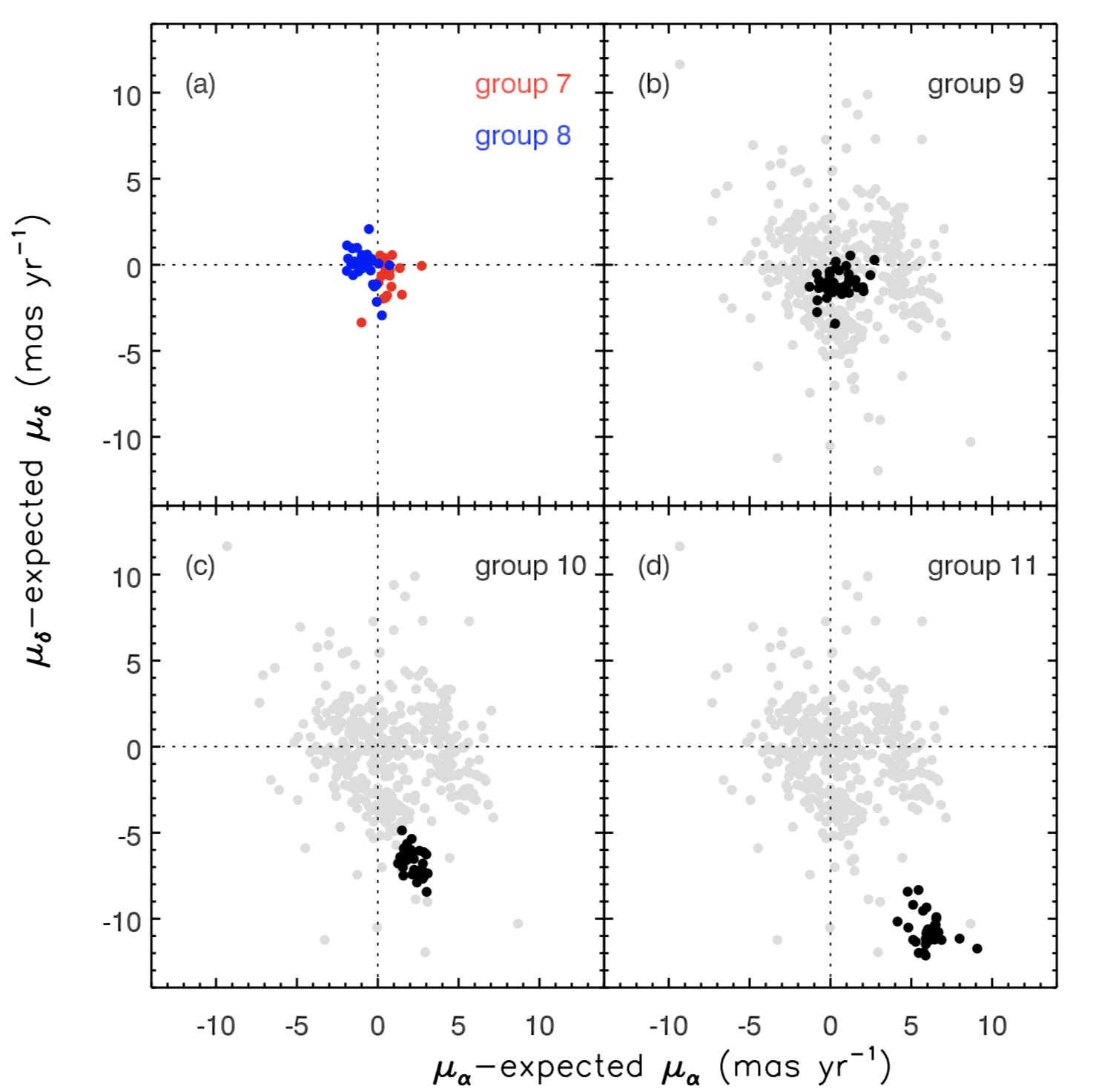}
\caption{(a) The residual proper motion diagram for Groups 7-8. The residual proper motion for each source is derived by subtracting its measured proper motions with the expected proper motions at its location but with the median space velocities of both groups. 
(b, c, d) The residual proper motion diagrams for identified YSOs (gray filled circles) and Groups 9-11 (black filled circles).  The residual proper motions for individual sources are derived by subtracting their measured proper motions with the expected proper motions at their locations but with the median UVW space velocities of the identified YSOs in Taurus.}
\label{pecy_young}
\end{figure}

In Group~7, the 6 known members belong to Population cyan in \cite{Luh2018} or Group C in \cite{Roc2020}. Group~7 locates to the east of Group~8, is at the similar distance and age to this group. Group~8 is also in Population cyan in \cite{Luh2018} or Group C in \cite{Roc2020}. Thus, there is one possibility that both Groups~7 and 8 belong to the same group and  their separation could be due to that we have not corrected for the projection effect in our grouping procedure. We search for the radial velocities (RVs) for the members in Groups~7 and 8 in {\em Gaia} DR2, and find the values for 3 sources. We correct for the project effect for Groups 7 and 8 employing the method in \cite{Kra2017}.  We derive the median values of the UVW velocities for the 3 sources and assume that all the members in the two groups share these median values in the UVW velocity space. For a group member, we subtract its measured proper motions with the ones expected at its location. Figure~\ref{pecy_young} (a) shows the residual proper motions for Groups 7 and 8, and the standard deviations of the residual proper motions for a combination of the two groups are both $\sim$1 mas~yr$^{-1}$, which are similar to the values of other young groups in the Taurus. Thus, it is very likely that both Groups~7 and 8 belong to the same group.

We further verify our grouping results for the young stars with the results in the literature. Compared with the ones in \cite{Luh2018},  Groups 1 and 5 belong to the red Population in \cite{Luh2018},  Groups 2, 3, 4 are in the blue Population, and Groups 6, 8 and  7 in the cyan Population. In order to compare with the result in \cite{Roc2020}, we cross-match the members of our young groups with the sources which have a probability of higher than 80\% of belonging to one of the Groups A-F in \cite{Roc2020}. We find that Group 1 in this work corresponds to Group A, Group 5 to Group B, Group 6 to Group D, Group 4 to Group E, and Group 2 to Group F, and both Groups 7 and 8 to Group C.

\begin{figure}
\centering
\includegraphics[angle=0,width=1\columnwidth]{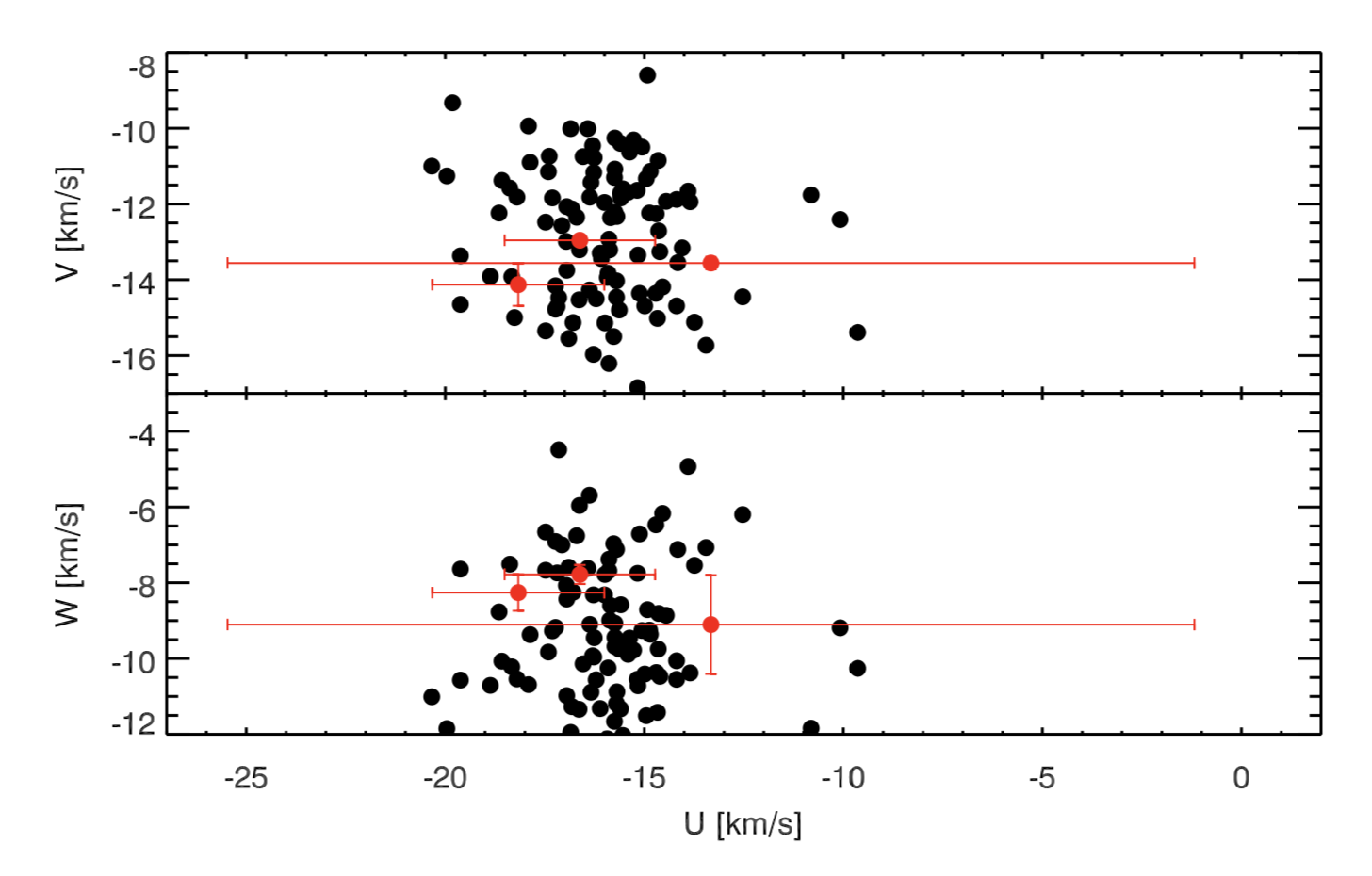}
\caption{Comparisons of the UVW velocities of three members (red filled circles ) in Group 9 with those of the identified YSOs (black solid dots, taken from \cite{Luh2018}). The typical errors of the UVW velocities for the YSOs are 0.23, 0.18 and 0.12\,\kms respectively. The uncertainties in the UVW velocities of the members in Group 9 are shown as red error bars.}
\label{UVW}
\end{figure}

\subsection{Old Groups}\label{Sect:Intermediate-age-Groups}

In this work, we find 14 groups, Groups 9-22, with ages ranging from 8 to 49~Myr and distances within $\sim$110--210\,pc.  The distributions of these groups are shown in Figure~\ref{location}. In these groups, there are 353 members and 37 have been catalogued in \cite{Kra2017}. In Group 11, there are 33 members and 6 of them are in the 118~Tau group (total 12 members) in \cite{Gag2018}. The age of 118 Tau is estimated to be $\sim$10\,Myr in \cite{Gag2018}, which is consistent with the age of Group 11. Therefore, it is likely that both Groups 11 and 118~Tau are from the same group.

\cite{Oh2017} searched the comoving star pairs by applying a marginalized likelihood ratio test to 3D velocities of stars of {\em Gaia} DR1 (e.g. TGAS: Tycho-Gaia Astrometric Solution). And they introduced 13,085 comoving star pairs among 10,606 unique stars. By comparing with their catalogue, we notice that about 20 members of the older groups of this work are also catalogued in \cite{Oh2017}. In Group 12 (total 52 members), there are 6 sources belonging to the star pair 29 (9 members) in \cite{Oh2017}, in Group 14 (total 30 members), 7 sources are in star pair 28 (9 members), and in another four old groups there are 1-3 sources which have been included in the star pairs in \cite{Oh2017}, see Table \ref{Tab:Tab:aggt_info} for more details. Excluding the ones overlapped with those in the literature, 7 old groups are discovered in this work, including Groups 9, 15, 16, 18, 19, 20, and 21.



We investigate the kinematic relations among the identified YSOs in Taurus and old groups in this work. Similar as done for Groups 7-8, we derive the residual proper motions for Groups 9--22. We adopt the UVW velocities for the identified YSOs from \cite{Luh2018}, and take the median values of them as the common values for the YSOs in Taurus. We derive residual proper motions for the old groups. For Groups 12-22,  their residual proper motions are far away from those known Taurus YSOs, indicating that they are not kinematically related to the known Taurus members. Furthermore, these groups are old ($>$20\,Myr) and the Taurus molecular clouds may not be present for so long \citep{Har2001,Har2012}. In Figure \ref{pecy_young} (b, c, d), we show the residual proper motions for the three relative younger groups (Groups 9--11), compared with the identified YSOs. For Group~9, the distribution of its residual proper motions are overlapped with those of known Taurus members, suggesting that there is a kinematic relation between them. In Group 9, there are 4 members (Source 290, 303, 312, 316) with the RVs in {\it Gaia} DR2. The RV of Star 316 significantly deviates from the other 3 ones, and is thus not used in this work. In Figure \ref{UVW} we compare the UVW velocities of the 3 sources in Group 9 with those of YSOs in Taurus, which indicates they are kinematically correlated. For Groups~10 and 11, the distributions of their residual proper motions are separated from the YSOs, which excludes that they have any kinematic relations with these YSOs.

\section{Summary}
In this work, we apply DBSCAN algorithm to the astrometric data of {\em Gaia} DR2  to search for groups in the Taurus field, and find 22 groups. We derive the ages of these groups by fitting them in color-magnitude diagram with the  isochrones of PARSEC models. According to their isochrone ages, the identified groups are divided into two categories: 8 young groups (2--4\,Myr), and 14 old groups (8--49\,Myr). A summary of the results in this work are listed as follows:

\begin{itemize}
\item[1.]
Among the young groups, we discover  17 new members. These are newly revealed young stars in Taurus star forming region.

\item[2.]
Among the 14 old groups, 7 ones, including Groups 9, 15, 16, 18, 19, 20, and 21, are firstly discovered in this work. 
\item[3.] 
Using infrared data from 2MASS and WISE, we characterize the disk properties of the sources in the 22 groups, confirm 104 disk-bearing stars in the literature and discover 19 new ones, 8 of which are in the young groups and 11 are in the 8--11\,Myr old groups.
\item[4.]
We use the strengths of H$\alpha$ emission line to characterize the accretion properties of the group members and discover one star (Source 332) with the accretion activity in 10\,Myr old group (Group 10).
\item[5.]
We find  a kinematic relation between Group~9 and the known Taurus members and exclude the relation between Groups~10-22 and the known Taurus members. 

\end{itemize}

\acknowledgments
We sincerely appreciate the anonymous referee for the instructive advice and productive suggestions.
This work is supported by National Key R\&D Program of China No. 2019YFA0405501. This work is supported by the National Natural Science Foundation of China (NSFC) with grants No. 11835057 to C. L. and 12003045 to J.M.L.. Guoshoujing Telescope (the Large Sky Area Multi-Object Fiber Spectroscopic Telescope LAMOST) is a National Major Scientific Project built by the Chinese Academy of Sciences. Funding for the project has been provided by the National Development and Reform Commission. LAMOST is operated and managed by the National Astronomical Observatories, Chinese Academy of Sciences. This work has made use of data from the European Space Agency (ESA) mission {\it Gaia} (\url{https://www.cosmos.esa.int/gaia}), processed by the {\it Gaia} Data Processing and Analysis Consortium (DPAC, \url{https://www.cosmos.esa.int/web/gaia/dpac/consortium}). Funding for the DPAC has been provided by national institutions, in particular the institutions participating in the {\it Gaia} Multilateral Agreement. Quite a substantial data processing of this work are executed through the software of TOPCAT \citep{Tay2005}.
\startlongtable
\begin{deluxetable*}{cccccccccccc}
\renewcommand{\tabcolsep}{0.05cm}
\tablecaption{A list of members in each group\label{Tab:star_inform}}
\tablehead{
\colhead{Group}&\colhead{ID$^{a}$}&\colhead{Name}&\colhead{RAJ2000}&\colhead{DEJ2000}&\colhead{Distance}& 
\colhead{SpT}&\colhead{Ref$^{b}$}&\colhead{New$^{c}$}&\colhead{Disk$^{d}$}&\colhead{C/W$^{e}$}&\colhead{$A_{J}$}\\
\colhead{}&\colhead{}&\colhead{}&\colhead{(deg)}&\colhead{(deg)}&\colhead{(pc)}&\colhead{}&\colhead{}
&\colhead{}&\colhead{}&\colhead{}&\colhead{(mag)}} 
\startdata
\hline
       1&       1&** KON    2A                   & 67.17759& 27.23450&132.4&M5.2&5& &5&    10.9,W& 0.06\\
       1&       2&EM* LkCa    4                  & 64.11711& 28.12659&129.4&M2  &5& &         &     3.4,W& 0.09\\
       1&       3&EM* LkCa   21                  & 65.51311& 28.42753&119.9&M2.5&5& &         &     5.5,W& 0.07\\
       1&       4&Gaia DR2 152416436443721728    & 65.28897& 27.84350&117.9&M5.2&5& &5&    24.8,C& 0.13\\
       1&       5&HD  28354                      & 67.33263& 27.40422&136.7&B9  &5& &         && 0.00\\
       1&       6&HD 283572                      & 65.49520& 28.30181&129.8&G4  &5& &         && 0.12\\
       1&       7&IRAS 04171+2756                & 65.10861& 28.06918&127.0&M3.5&5& &5&     3.9,W& 0.06\\
       1&       8&IRAS F04147+2822               & 64.45691& 28.49342&128.6&M3.7&5& &5&     6.4,W& 0.41\\
\enddata
\label{Tab:star}
\centering
\tablenotetext{a\,}{The identification numbers of the stars in this work.}
\tablenotetext{b\,}{The references for the spectral types:  1. \cite{Mul1950}; 2. \cite{Hei1975}; 3. \cite{Sle2006}; 4. \cite{Dunkin1998}; 5. \cite{Esp2019}; 6. from the LAMOST catalog; 7. This work; 8. \cite{Can1993}; 9. \cite{Nes1995}; 10. \cite{Bia2012}; 11. \cite{Her2014}; 12. \cite{Gag2018b}; 13. \cite{Ker2019}}
\tablenotetext{c\,}{New members (``Y'') or nonmembers (``N'') of the young groups (2--4\,Myr) identified in this work.}
\tablenotetext{d\,}{When the sources are identified as having circumstellar disks, the numbers are for the references : 7. This work; 14. \cite{Mah2002}; 15. \cite{Jen2003}; 16. \cite{Eis2004}; 17. \cite{Reb2011}; 18. \cite{Men2015}}
\tablenotetext{e\,}{``C'' denotes CTTS and  ``W'' for WTTS, the value in front of ``C'' or ``W'' is the H$\alpha$ $EWs$ derived from LAMOST spectra.}
\tablenotetext{f\,}{The 20 disk bearing stars of \cite{Esp2019} that not confirmed in this work.}
\end{deluxetable*}

\begin{table*}
\caption{ A list of the groups identified in this work.}
\scriptsize
\centering
\label{Tab:Tab:aggt_info}
\begin{tabular}{cccccccccccc}

\hline\hline
group  & members&PK$^{a}$&Other name$^{b}$&RAJ2000$^{c}$ & DEJ2000$^{c}$  & pmRA$^{c}$ & pmDEC$^{c}$ & Distance$^{c}$ & Age$^d$ & Age$^e$ &A$_{J}^{c}$ \\

&&&&(deg)&(deg)&(mas/yr)&(mas/yr)&(pc)&(Myr)&(Myr)&(mag)\\
\hline
       1&55&          55&A& 65.22& 27.91&  8.68&-25&129$ \pm^{7}_{12}$ &3&3&  0.21\\
\hline       
       2&41&          40&F& 68.88& 17.64& 12.15&-18&144$ \pm^{8}_{8}$ &3&2&  0.12\\
\hline       
       3&19&          19&& 66.13& 26.65& 11.16&-17&159$ \pm^{10}_{8}$ &3&3&  0.57\\
\hline       
       4&27&          26&E& 68.69& 22.88&  9.89&-17&158$ \pm^{8}_{10}$ &2&2&  0.45\\
\hline       
       5&49&          49&B& 68.53& 24.79&  7.15&-21&130$ \pm^{13}_{10}$ &2&2&  0.25\\
\hline       
       6&7&           5&D& 76.88& 25.02&  2.78&-17&172$ \pm^{5}_{4}$ &4&6&  0.14\\
\hline\hline
\end{tabular}

\tablenotetext{a\,}{Numbers of the stars with known spectral types in the literature \citep{Sle2006,Kra2017,Esp2019}.}
\tablenotetext{b\,}{``A''-``F'' denote the Groups of \cite{Roc2020}. Oh17 and ``Oh17-X(c/t)'' indicate \cite{Oh2017} and the star pair of it. ``X'' denote the name of the star pair, ``t'' denotes the total members of the star pair, while ``c'' denotes how many members contains in the catalogue of this work.}
\tablenotetext{c\,}{The mean position, distance, proper motion and $J$-band extinction of each individual group}
\tablenotetext{d\,}{The ages of the groups that derived by fitting the dereddened color-magnitude diagrams with the isochrones of PARSEC models. The extinction of each individual star is considered.}
\tablenotetext{e\,}{The ages derived by fitting the observed color-magnitude diagrams with the isochrones of PARSEC models, but assuming that all members of a group share the same extinction and treating the extinction as a free parameter.}
\end{table*}

\end{document}